\documentclass{aastex}
\usepackage{emulateapj5}
\usepackage{epsfig}
\renewcommand{\placetable}[1]{}
\renewcommand{\placefigure}[1]{}
\renewcommand{\tablecolumns}[1]{}
\renewcommand{\tablewidth}[1]{}
\renewcommand{\tabletypesize}[1]{#1}
\renewcommand{\colhead}[1]{#1}
\renewcommand{\startdata}{\\ \hline \noalign{\smallskip}}
\renewcommand{\enddata}{\hline \end{tabular}\smallskip\\}
\renewcommand{\tablenotetext}[2]{\footnotetext{}{\smallskip $^{\rm #1}$#2}\\}
\renewcommand{\tablerefs}[1]{References. --- #1}
\renewcommand{\tablecomments}[1]{Note. --- #1\smallskip\\}
\renewcommand{\sqrt}[1]{{#1}^{1/2}}
\newcommand{\wcap}{0.476\textwidth}
\slugcomment{The Astrophysical Journal, 564:\#\#\#--\#\#\#, 2002 January 10}

\newcommand{\Berat}{$^{10}$Be/$^9$Be}
\newcommand{\gray}{$\gamma$-ray}
\newcommand{\grays}{$\gamma$-rays}

\newcommand{\Dpp}{D_{pp}}
\newcommand{\Dxx}{D_{xx}}
\newcommand{\ddp}{\frac{\partial}{\partial p}}

\newcommand{\hi}{H {\sc i}}
\newcommand{\hii}{H {\sc ii}}
\newcommand{\twofigs}{0.49\textwidth}
\newcommand{\onefig}{0.95\textwidth}
\newcommand{\adv}{Adv.\ Space Res.}
\newcommand{\app}{Astropart.\ Phys.}

\newcommand{\plb}{Phys.\ Lett.\ B}
\newcommand{\nphysb}{Nucl.\ Phys.\ B}
\newcommand{\pub}[4]{#4, #1, #2, #3}

\shorttitle{Secondary antiprotons and propagation of cosmic rays}
\shortauthors{Moskalenko et al.}
\begin{document}
\title{Secondary antiprotons and propagation of cosmic rays in the Galaxy and heliosphere}

\author{Igor V.~Moskalenko\altaffilmark{1,2}}
\affil{NASA/Goddard Space Flight Center, Code 660, Greenbelt, MD 20771}
\altaffiltext{1}{NAS/NRC Senior Research Associate}
\altaffiltext{2}{Also Institute for Nuclear Physics, 
   M.V.Lomonosov Moscow State University, 119 899 Moscow, Russia}
\email{imos@milkyway.gsfc.nasa.gov}

\author{Andrew W.~Strong}
\affil{Max-Planck-Institut f\"ur extraterrestrische Physik,
   Postfach 1603, D-85740 Garching, Germany}
\email{aws@mpe.mpg.de}

\author{Jonathan F.~Ormes} 
\affil{NASA/Goddard Space Flight Center, Code 600, Greenbelt, MD 20771}
\email{jfo@lheapop.gsfc.nasa.gov}


\author{Marius S.~Potgieter}
\affil{Unit for Space Physics, Potchefstroom University for CHE, 
2520 Potchefstroom, South Africa}
\email{fskmsp@puknet.puk.ac.za}
\bigskip
\submitted{Received 2001 June 20; accepted 2001 September 17 \bigskip}
\begin{abstract}

High-energy collisions of cosmic-ray nuclei with interstellar gas are
believed to be the mechanism producing the majority of cosmic ray
antiprotons. Due to the kinematics of the process they are created
with a nonzero momentum; the characteristic spectral shape with a
maximum at $\sim2$ GeV and a sharp decrease towards lower energies
makes antiprotons a unique probe of models for particle propagation in
the Galaxy and modulation in the heliosphere. On the other hand,
accurate calculation of the secondary antiproton flux provides a
``background'' for searches for exotic signals from the annihilation
of supersymmetric particles and primordial black hole evaporation.
Recently new data with large statistics on both low and high energy
antiproton fluxes have become available which allow such tests to be
performed. We use our propagation code GALPROP to calculate
interstellar cosmic-ray propagation for a variety of models. We show
that there is \emph{no} simple model capable of accurately describing
the whole variety of data: boron/carbon and sub-iron/iron ratios,
spectra of protons, helium, antiprotons, positrons, electrons, and
diffuse \grays. We find that only a model with a break in the
diffusion coefficient plus convection can reproduce measurements of
cosmic-ray species, and the reproduction of primaries ($p$, He) can be
further improved by introducing a break in the primary injection
spectra. For our best-fit model we make predictions of proton and
antiproton fluxes near the Earth for different modulation levels and
magnetic polarity using a steady-state drift model of propagation in
the heliosphere.
\end{abstract}



\section{Introduction}
\label{sec:intro}


\begin{table*}[!th]
\tablecolumns{6}
\tablewidth{0mm}
\tabletypesize{\footnotesize}

\caption{Summary of recent CR antiproton calculations. \label{table1}}
\begin{tabular}{lllcccc}\hline\hline\noalign{\smallskip}

\colhead{} & 
\colhead{Antiproton} & 
\colhead{Ambient CR} & 
\colhead{Propagation} & 
\colhead{Reacceleration/} &
\colhead{Modulation} & 
\colhead{Tertiary}
\\
\colhead{Model} & 
\colhead{production} & 
\colhead{spectrum} & 
\colhead{model} & 
\colhead{Convection} & 
\colhead{in heliosphere} & 
\colhead{antiprotons}
\startdata
\citet*{berg99}  &
Tan \& Ng        & 
scaled LIS       &
simplified       & 
n/y              & 
force-field      & 
y                \\
 &
 & 
 &
diffusion        \smallskip\\

\citet{bieber99} & 
DTUNUC           & 
LIS, in the      &
leaky-box        & 
n/n              & 
steady-state     & 
y                \\
 &
event generator  &
whole Galaxy     &
 &
 &
drift model\tablenotemark{a}\smallskip\\

\citet{bottino98}& 
Tan \& Ng        & 
LIS, in the      &
two zone         & 
n/n              & 
force-field      & 
n                \\
 & 
 &
whole Galaxy     &
diffusion        \smallskip\\

\citet{donato01} & 
Tan \& Ng,       & 
LIS, in the      &
two zone         & 
y/y              & 
force-field      & 
y                \\
 &
DTUNUC           &
whole Galaxy     & 
diffusion        \smallskip\\

\citet*{simon98} & 
DTUNUC           & 
LIS, in the      &
leaky-box        & 
n/n              & 
force-field      & 
n                \\
 &
event generator  &
whole Galaxy     \medskip\\

{\it Present work}&
$pp$: Tan \& Ng  &
propagated       &
3D spatial &
y/y              &
steady-state     &
y                \\
 &
$pA$: DTUNUC     &
 & 
grid, realistic  &
 & 
drift model      \\
 & 
 &
 & 
diffusion        \\ 
\enddata
\tablenotetext{a}{Parameters are fixed from the assumed LIS proton spectrum.}

\end{table*}

Most of the CR antiprotons observed near the Earth are secondaries
produced in collisions of energetic CR particles with interstellar gas
\citep[e.g.,][]{mitchell}. Due to the kinematics of this process, the
spectrum of antiprotons has a unique shape distinguishing it from
other cosmic-ray species. It peaks at about 2 GeV decreasing sharply
towards lower energies. In addition to secondary antiprotons there
are possible sources of primary antiprotons; those most often
discussed are the dark matter particle annihilation and evaporation of
primordial black holes (PBHs).

The nature and properties of the dark matter that constitute a
significant fraction of the mass of the Universe have puzzled
scientists for more than a decade \citep{trimble,ashman}. Among the
favored dark matter candidates are so-called weakly-interacting
massive particles (WIMPs), whose existence follows from
supersymmetric models \citep*[for a review see][]{jkg}. Such
particles, if stable, could have a significant cosmological abundance
and be present in our own Galaxy. A pair of stable WIMPs can
annihilate into known particles and antiparticles making it possible
to infer WIMPs in the Galactic halo by the products of their
annihilations. PBHs may have formed in the early Universe via initial
density fluctuations, phase transitions, or the collapse of cosmic
strings \citep*{hawking,carr,maki}. Black holes can emit particles and
evaporate due to quantum effects. The emission rate is generally too
low to be observable, but it increases as the black hole mass
decreases. The only observable PBHs are those that have a mass small
enough to produce a burst of particles as they evaporate.

In recent years, new data with large statistics on both low and high
energy antiproton fluxes have become available
\citep{hof,basini,Orito00,bergstroem00,Maeno01,Stoc01,asaoka01} that allow us
to test models of CR propagation and heliospheric modulation. A probe
to measure low energy particles in interstellar space may also become
reality in the near future \citep{probe}. Additionally, accurate
calculation of the secondary antiproton flux provides a ``background''
for searches for exotic signals such as WIMP annihilation or PBH
evaporation.

Despite numerous efforts and overall agreement on the secondary nature
of the majority of CR antiprotons, published estimates of the expected
flux significantly differ \citep[see, e.g., Fig.\ 3 in][]{Orito00}.
Calculation of the secondary antiproton flux is a complicated task.
The major sources of uncertainties are three fold: (i) incomplete
knowledge of cross sections for antiproton production, annihilation,
and scattering, (ii) parameters and models of particle propagation in
the Galaxy, and (iii) modulation in the heliosphere. While the
interstellar antiproton flux is affected only by uncertainties in the
cross sections and propagation models, the final comparison with
experiment can only be made after correcting for the solar
modulation. Besides, the spectra of CR nucleons have been directly
measured only inside the heliosphere while we need to know the
spectrum outside, in interstellar space, to compute the antiproton
production rate correctly. The basic features of the recent
models are summarized in Table \ref{table1}.

\placetable{table1}

For the antiproton production cross section two options exist: a
semiphenomenological fit to the data by \citet{TanNg83a,TanNg83b}
which has been used for almost two decades because of lack of new data
and the Monte Carlo event generator DTUNUC
\citep*{dtunuc1,dtunuc2}. Both the parametrization and the DTUNUC code
describe well the available data on antiproton production in
$pp$-collisions. For interactions involving heavier nuclei there are
no measurements so far. However, the DTUNUC model provides a
reasonable description of hadron-nucleus and nucleus-nucleus
interactions and thus it can be used for estimates of antiproton
production in proton-nucleus collisions. Additionally, ``tertiary''
antiprotons, inelastically scattered secondaries, are the main
component at low energies. This component has only been included in
the most recent models.

To calculate particle propagation one must choose between the
easy-to-apply but non-physical ``leaky-box'' model, and a variety of
diffusion models. The leaky-box model reduces the problem to
derivation of the path length distribution \citep*[which in turn remains the
major source of uncertainty as discussed, e.g., by][]{simon98}.
Diffusion models are theoretically more physical, but
also differ by the degree to which they reflect reality. The ultimate
goal is to develop a model which is consistent with many different
kinds of data available: measurements of CR species, \grays, and
synchrotron emission.

Heliospheric modulation is not yet understood in detail. The first and
yet popular force-field approximation \citep{forcefield} does not work
at low energies. More sophisticated drift models reflect our still
incomplete current knowledge. The main problem here is that the
modulation parameters are determined based on the assumed {\it ad hoc}
interstellar nucleon spectrum. The antiproton/proton ratio, which is
often calculated, is even more uncertain at low energies and may vary
by an order of magnitude over the solar cycle
\citep{LabradorMewaldt97}. It is therefore the antiproton spectrum
itself which should be compared with calculations. On the other hand,
a reliable calculation of the antiproton spectrum would allow the
study of charge sign dependent effects in the heliosphere. That
could, in turn, help to derive the low-energy part of the local
interstellar (LIS) spectrum of nucleons and help to establish the
heliospheric diffusion coefficients more accurately.

We have developed a numerical method and corresponding computer code
GALPROP\footnote{Our model including software and datasets is
available at
\url{http://www.gamma.mpe--garching.mpg.de/$\sim$aws/aws.html}} for
the calculation of Galactic CR propagation in 3D \citep{SM98}. The
code has been shown to reproduce simultaneously observational data of
many kinds related to CR origin and propagation \citep[for a review
see][]{SM99,MS00}. The code has been validated on direct measurements
of nuclei, antiprotons, electrons, and positrons, and astronomical
measurements of \grays\ and synchrotron radiation. These data provide
many independent constraints on model parameters.

The code is sufficiently general that new physical effects can be
introduced as required. The basic spatial propagation mechanisms are
diffusion and convection, while in momentum space energy loss and
diffusive reacceleration are treated. Fragmentation, secondary
particle production, and energy losses are computed using realistic
distributions for the interstellar gas and radiation fields. Recently
the code has been entirely rewritten in C++ with many new features
added \citep{SM01}.

In this paper we use the GALPROP code for accurate calculation of
production and propagation of secondary antiprotons and positrons. We
explore the dependence of the antiproton and positron fluxes on the
nucleon injection spectrum and propagation parameters. The antiproton
production cross section is calculated using the $pp$ production cross
section \citep{TanNg83b} and DTUNUC nuclear factors from
\citet{simon98} to calculate $\bar p$ production in $pA$ collisions
or the $pp$ production cross section scaled
appropriately with beam/target atomic numbers. Inelastic scattering
to produce ``tertiary'' antiprotons and ``secondary'' protons is
taken into account. The calculated proton and antiproton LIS and
$\bar p/p$ ratio are modulated using the steady-state drift model.
Our preliminary results have been reported in \citet{moskalenko01}.

\section{Basic features of the GALPROP models} \label{sec:descr}

The cylindrically symmetric GALPROP models have been described in
detail elsewhere \citep{SM98}; here we summarize their basic features.

The models are three dimensional with cylindrical symmetry in the
Galaxy, and the basic coordinates are $(R,z,p)$ where $R$ is
Galactocentric radius, $z$ is the distance from the Galactic plane and
$p$ is the total particle momentum. In the models the propagation
region is bounded by $R=R_h$, $z=\pm z_h$ beyond which free escape is
assumed.

The propagation equation we use for all CR species is written in the form:
\begin{eqnarray}
\label{eq.1}
\frac{\partial \psi}{\partial t} 
&=& q({\mathbf r}, p)+ \nabla \cdot ( \Dxx\nabla\psi - {\mathbf V}\psi )
+ \ddp\, p^2 \Dpp \ddp\, \frac{1}{p^2}\, \psi \nonumber\\
&-& \frac{\partial}{\partial p} \left[\dot{p} \psi
- \frac{p}{3} \, (\nabla \cdot {\mathbf V} )\psi\right]
- \frac{1}{\tau_f}\psi - \frac{1}{\tau_r}\psi\, ,
\end{eqnarray}
where $\psi=\psi ({\mathbf r},p,t)$ is the density per unit of total
particle momentum, $\psi(p)dp = 4\pi p^2 f({\mathbf p})$ in terms of
phase-space density $f({\mathbf p})$, $q({\mathbf r}, p)$ is the source term,
$\Dxx$ is the spatial diffusion coefficient, ${\mathbf V}$ is the
convection velocity, reacceleration is described as diffusion in
momentum space and is determined by the coefficient $\Dpp$,
$\dot{p}\equiv dp/dt$ is the momentum loss rate, $\tau_f$ is the time
scale for fragmentation, and $\tau_r$ is the time scale for
radioactive decay. The numerical solution of the transport equation is
based on a Crank-Nicholson \citep{Press92} implicit second-order
scheme. The three spatial boundary conditions $\psi(R_h,z,p) =
\psi(R,\pm z_h,p) = 0$ are imposed on each iteration, where we take
$R_h=30$ kpc.

For a given $z_h$ the diffusion coefficient as a function of momentum
and the reacceleration or convection parameters is determined by
boron-to-carbon (B/C) ratio data. The spatial diffusion coefficient
is taken as $\Dxx = \beta D_0(\rho/\rho_0)^{\delta}$ if necessary with
a break ($\delta=\delta_1$ below rigidity $\rho_0$, $\delta=\delta_2$
above rigidity $\rho_0$). The injection spectrum of nucleons is
assumed to be a power law in momentum, $dq(p)/dp \propto p^{-\gamma}$
for the injected particle density.

Reacceleration provides a mechanism to reproduce the B/C ratio without
an ad-hoc form for the diffusion coefficient. Our reacceleration
treatment assumes a Kolmogorov spectrum with $\delta=1/3$ or value
close to this. For the case of reacceleration the momentum-space
diffusion coefficient $D_{pp}$ is related to the spatial coefficient
$\Dxx$ \citep{berezinskii,seo} via the Alfv\'en speed $v_A$.

The convection velocity (in $z$-direction only) $V(z)$ is assumed to
increase linearly with distance from the plane ($dV/dz>0$ for all
$z$). This implies a constant adiabatic energy loss; the possibility
of adiabatic energy gain ($dV/dz < 0$) is not considered. The linear
form for $V(z)$ is consistent with cosmic-ray driven MHD wind models
\citep[e.g.,][]{Zirakashvili96}. The velocity at $z = 0$ is a model
parameter, but we consider here only $V(0) = 0$.
($V(0) = 0$ is physically plausible: it is required to avoid a discontinuity,
since by symmetry a non-zero velocity would have to change sign at $z=0$.)

\centerline{\psfig{file=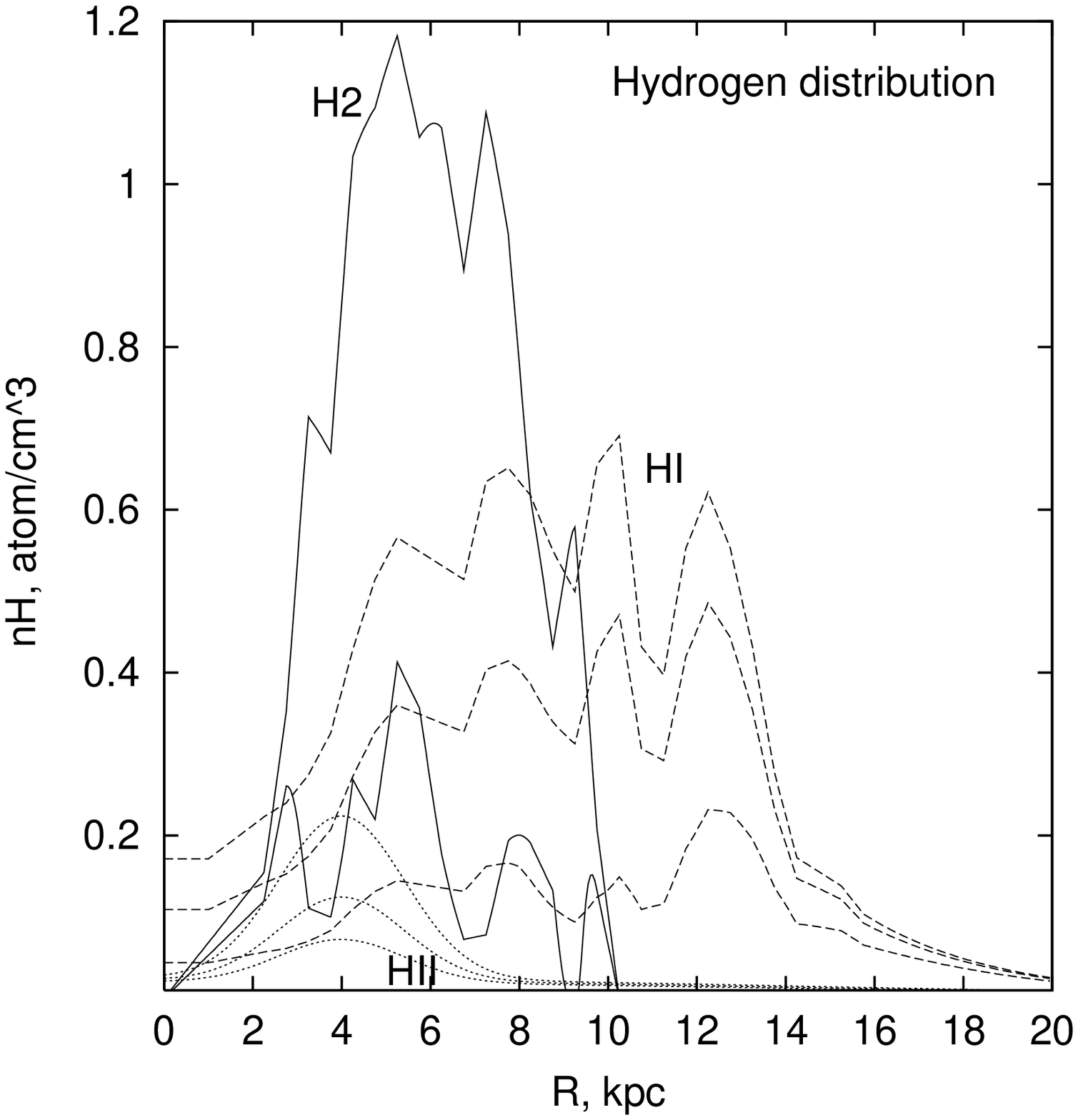,width=\twofigs,clip=}}
\figcaption[f1.ps]{ Number density distributions of $2\times$H$_2$
(solid), \hi\ (dashes), and \hii\ (dots) in the Galaxy. Shown are the
plots for $z=0, 0.1, 0.2$ kpc (decreasing density). Number density of
H$_2$ at $z=0.2$ kpc from the plane is very low and is not shown in
the plot.
\label{fig:hydrogen} \bigskip} 

The interstellar hydrogen distribution uses \hi\ and CO surveys
and information on the ionized component; the helium fraction of the
gas is taken as 0.11 by number. The H$_2$ and \hi\ gas number
densities in the Galactic plane are defined in the form of tables,
which are interpolated linearly. The extension of the gas distribution
to an arbitrary height above the plane is made using analytical
approximations. The distributions of $2\times$H$_2$, \hi, and
\hii\ are plotted on Figure \ref{fig:hydrogen} for $z =0, 0.1,
0.2$ kpc. The code uses the densities averaged over the $z$-grid
using a 0.01 kpc step. More details about the gas distribution model
are given in Appendix \ref{gas}.

\placefigure{fig:hydrogen}

The distribution of cosmic-ray sources is chosen to reproduce the
cosmic-ray distribution determined by analysis of EGRET \gray\ data
\citep{StrongMattox96} and was described in \citet{SM98}.

Energy losses for nucleons by ionization and Coulomb interactions are
included, and for electrons by ionization, Coulomb interactions,
bremsstrahlung, inverse Compton, and synchrotron.

Positrons and electrons (including secondary electrons) are propagated
in the same model. Positron production is computed as described in
\citet{MS98}, that paper includes a critical reevaluation of the
secondary $\pi^\pm$- and $K^\pm$-meson decay calculations.

Gas-related \gray\ intensities are computed from the emissivities as a
function of $(R,z,E_\gamma)$ using the column densities of \hi\ and
H$_2$. The interstellar radiation field (ISRF), used for calculation
of the inverse Compton (IC) emission and electron energy losses, is
calculated based on stellar population models and COBE results, plus
the cosmic microwave background.

An overview of our previous analyses is presented in \citet{SM99} and
\citet{MS00} and full results for protons, helium, positrons, and
electrons in \citet{MS98} and \citet*{SMR00}.
The evaluations of the B/C and \Berat\ ratios, diffusion/convection
and reacceleration models and full details of the numerical method are
given in \citet{SM98}. Antiprotons have been previously discussed in
the context of the ``hard interstellar nucleon spectrum'' hypothesis
in \citet*{MSR98} and \citet{SMR00}. Our results for diffuse
continuum \grays, synchrotron radiation, and a new evaluation of the
ISRF are described in \citet{SMR00}.

\subsection{New developments} \label{sec:new}

The experience gained from the original fortran--90 code allowed us to
design a new version of the model, entirely rewritten in C++, that is
much more flexible. It allows essential optimizations in comparison
to the older model and a full 3-dimensional spatial grid. It is now
possible to explicitly solve the full nuclear reaction network on a
spatially resolved grid. The code can thus serve as a complete
substitute for the conventional ``leaky-box'' or ``weighted-slab''
propagation models usually employed, giving many advantages such as
the correct treatment of radioactive nuclei, realistic gas and source
distributions etc. It also allows stochastic SNR sources to be
included. It still contains an option to switch to the fast running
cylindrically symmetrical model which is sufficient for many
applications such as the present one.

In the new version, we have updated the cross-section code to include
the latest measurements and energy dependent fitting functions. The
nuclear reaction network is built using the Nuclear Data Sheets. The
isotopic cross section database consists of more than 2000 points
collected from sources published in 1969--1999. This includes a
critical re-evaluation of some data and cross checks. The isotopic
cross sections are calculated using the author's fits to major
beryllium and boron production cross sections $p+\mathrm{C,N,O} \to \mathrm{
Be,B}$. Other cross sections are calculated using
\citet*{W-code}\footnote {Code \url{WNEWTR.FOR}, version 1993, as
posted at\\
\url{http://spdsch.phys.lsu.edu/SPDSCH\_Pages/Software\_Pages/Cross\_Section\_Calcs/WebberKishSchrier.html
} (some minor changes have been made to make it compatible with
GALPROP).} and/or \citet*{ST-code}\footnote {Code
\url{YIELDX\_011000.FOR}, version 2000, as posted at\\
\url{http://spdsch.phys.lsu.edu/SPDSCH\_Pages/Software\_Pages/Cross\_Section\_Calcs/SilberburgTsao.html}
(some minor changes have been made to make it compatible with
GALPROP).} phenomenological approximations renormalized to the data
where it exists. The cross sections on the He target are calculated
using a parametrization by \citet{ferrando88}.

The reaction network is solved starting at the heaviest nuclei (i.e.,
$^{64}$Ni). The propagation equation is solved, computing all the
resulting secondary source functions, and then proceeds to the nuclei
with $A-1$. The procedure is repeated down to $A=1$. In this way all
secondary, tertiary etc.\ reactions are automatically accounted for.
This includes secondary protons, inelastically scattered primaries;
their energy distribution after scattering is assumed to be the same
as for antiprotons (eq.\ [\ref{eq.31}]). To be completely accurate
for all isotopes, e.g., for some rare cases of $\beta^\pm$-decay, the
whole loop is repeated twice. Our preliminary results for all cosmic
ray species $Z\leq28$ are given in \citet{SM01}.

For the calculation reported here, we use a cylindrically symmetrical
Galactic geometry.

\subsection{Antiproton cross sections} \label{sec:cs}

We calculate $\bar p$ production and propagation using the basic
formalism described in \citet{MSR98}. Antiproton production in
$pp$-collisions has been calculated using the parametrization of the
invariant $\bar p$-production cross section given by \citet{TanNg83b}.
This parametrization fits available data quite well. It gives an
antiproton multiplicity slightly higher than the DTUNUC code just
above the threshold and agrees with DTUNUC results at higher energies
\citep{simon98}.

For the cross sections $\sigma_{pp}^{inel}$ and $\sigma_{pA}^{inel}$
we adapted parametrizations by \citet{TanNg83a}, \citet{PDG}, and
\citet*{Letaw83}. At low energies the total $\bar
pp$ inelastic cross section has been calculated using a fit from
\citet{TanNg83a}. At high energies it is calculated as the difference
between total $\bar pp$ and $pp$ cross sections which are parametrized
using Regge theory \citep{PDG}. The $\bar p$ absorption cross section
on an arbitrary nuclear target has been scaled by $A^{2/3}$ using the
measured $\bar p$--C,Al,Cu cross sections \citep{pbar_A}.

To this we have added $\bar p$ annihilation and treated inelastically
scattered $\bar p$'s as a separate ``tertiary'' component. The energy
distribution after scattering is assumed to be \citep{TanNg83a}
\begin{equation}
\label{eq.31}
\frac{dN(E_{\bar p},E_{\bar p}')}{dE_{\bar p}} = \frac{1}{T_{\bar p}'},
\end{equation}
where $E_{\bar p}'$ and $E_{\bar p}$ are the total $\bar p$ energy
before and after scattering correspondingly, and $T_{\bar p}'$ is the
$\bar p$ kinetic energy before scattering.

The $\bar p$ production by nuclei with $Z\geq2$ is calculated using
effective nuclear factors by \citet{simon98} and scaling factors
similar to \citet{GaisserSchaefer92}.

In the first method we use the effective factor obtained from
simulations of the $\bar p$ production with the Monte Carlo model
DTUNUC, which appear to be more accurate than simple scaling. The use
of this factor is consistent since the proton spectrum adapted in
\citet{simon98} is close to our propagated spectrum above the $\bar p$
production threshold. For convenience, we made a fit to the ratio of
the total $\bar p$ yield to the $\bar p$ yield from the $pp$-reaction
\citep[column 3/column 2 ratio as given in Table 2 in][]{simon98}:
\begin{equation}
\label{eq.32}
\sigma_\Sigma/\sigma_{pp} = 0.12\,(T_{\bar p}/\mathrm{GeV})^{-1.67}+1.78,
\end{equation}
where $T_{\bar p}$ is the kinetic $\bar p$ energy.

In the second method the cross section for $\bar p$ production in
proton-nucleus and nucleus-nucleus interactions has been obtained by
scaling the $pp$ invariant cross section with a factor
\citep{GaisserSchaefer92}
\begin{equation}
\label{eq.33}
F_{it\to\bar pX}= 1.2 (A_i \sigma_{pt}^{inel}+A_t \sigma_{pi}^{inel})
/2\sigma_{pp}^{inel},
\end{equation}
where $A_{i,t}$ are the atomic numbers of the incident and target
nuclei, $\sigma_{pt}, \sigma_{pi}$ are the $pA_{i,t}$ cross sections,
and a factor 1.2 is put for consistency with \citet{simon98}
calculations. Production of low-energy antiprotons in $p\alpha$- and
$\alpha\alpha$-reactions is increased over that for simple scaling.
In order to account for this effect we put $\tilde E_{\bar p} =
E_{\bar p}+0.06$ GeV instead of $E_{\bar p}$ when calculating the
source spectra.

Detailed results published by \citet{simon98} (their Figs.\ 2, 4, 7)
allow us to test the approximation given by equation (\ref{eq.33}) and show its
equivalence to DTUNUC calculations and to equation (\ref{eq.32}). For
these tests we used interstellar proton and helium spectra as
published by \citet{Menn00}; these spectra were used by
\citet{simon98}. The main discrepancy is $\sim15-20$\%
under-production of antiprotons in $pp$-collisions between $T_{\bar
p}=5-30$ GeV compared to Simon et al.\ calculations (their Fig.\ 7),
but this does not influence production and propagation of antiprotons
at lower energies. At energies below $\sim5$ GeV the factors given by
equation (\ref{eq.32}) and equation (\ref{eq.33}) are virtually equivalent.
We further use equation (\ref{eq.32}) in our calculations.

The $\bar pp$ elastic scattering is not included. At sufficiently high
energies it is dominated by the forward peak with small energy
transfer while at low energies the inelastic cross section (mostly
annihilation) accounts for about 70\% of the total cross section
\citep{elastic1,elastic2}.

As we will see, the accuracy of antiproton cross sections is now a
limiting factor; we therefore summarize the current status in
Appendix \ref{sec:appendix}.

\section{Solar modulation} \label{sec:modulation}

Thanks to the Ulysses mission significant progress has been made in
our understanding of the major mechanisms driving the modulation of CR
in the heliosphere. The principal factors in modulation modeling are
the heliospheric magnetic field (HMF), the solar wind speed, the tilt
of the heliospheric current sheet, and the diffusion tensor. Here, we
give a short description of the model used in this paper for
modulation of galactic protons and antiprotons. More detail on the
characteristics of heliospheric modulation and modulation models can
be found elsewhere \citep[e.g.,][and references therein]{modulation,Burger}.

Modulation models are based on the numerical solution of the cosmic
ray transport equation \citep{Parker65}:
\begin{eqnarray}
\label{eq.40}
\frac{\partial f({\mathbf r},\rho,t)}{\partial t}
=&-&({\mathbf V}+\langle{\mathbf v}_D\rangle)\cdot \nabla f \nonumber\\
&+&\nabla \cdot ({\mathbf K}_S\cdot\nabla f)
+\frac{1}{3} (\nabla\cdot {\mathbf V})\frac{\partial f}{\partial\ln\rho},
\end{eqnarray}
where $f({\mathbf r},\rho,t)$ is the CR distribution function,
$\mathbf{r}$ is the position, $\rho$ is the rigidity, and $t$ is
time. Terms on the right-hand side represent convection, gradient and
curvature drifts, diffusion, and adiabatic energy losses respectively,
with $\mathbf{V}$ the solar wind velocity. The symmetric part of the
tensor ${\mathbf K}_S$ consists of diffusion coefficients parallel
($K_\parallel$) and perpendicular ($K_{\perp\theta}$ and $K_{\perp
r}$) to the average HMF. The anti-symmetric element $K_A$ describes
gradient and curvature drifts in the large scale HMF where the pitch
angle averaged guiding center drift velocity for a near isotropic CR
distribution is given by $\langle{\mathbf v}_D\rangle=\nabla\times K_A
{\mathbf e}_B$, with ${\mathbf e}_B={\mathbf B}/B$, and $B$ the
magnitude of the background magnetic field. The effective radial
diffusion coefficient is given by $K_{rr}= K_\parallel \cos^2\psi
+K_{\perp r} \sin^2\psi$, with $\psi$ the angle between the radial
direction and the averaged magnetic field direction. We enhanced
perpendicular diffusion in the polar direction by assuming
$K_{\perp\theta} > K_{\perp r}$ in the heliospheric polar regions
\citep[see also][]{Potgieter00}.

Drift models predict a clear charge-sign dependence for the
heliospheric modulation of positively charged particles, e.g., CR
protons and positrons, and negatively charged particles, e.g.,
electrons and antiprotons. This is due to the different large-scale
gradient, curvature and current sheet drifts that charged particles
experience in the HMF. For example, antiprotons will drift inwards
primarily through the polar regions of the heliosphere during $A<0$
polarity cycles, when the HMF is directed towards the Sun in
the northern hemisphere. Protons, on the other hand, will then drift
inwards primarily through the equatorial regions of the heliosphere,
encountering the wavy heliospheric current sheet in the
process. During the $A > 0$ polarity cycles the drift directions for
the two species reverse, so that a clear 22-year cycle is caused
\citep[e.g.,][]{cycle22}.

We use a steady-state two-dimensional model that simulates the effect
of a wavy current sheet \citep{current_sheet} by using an averaged
drift field with only $r$- and $\theta$-components for the
three-dimensional drift pattern in the region swept out by the wavy
current sheet. The solar wind speed is 400 km/s in the equatorial
plane and increases to 800 km/s in the polar regions
\citep*[for details see][]{Burger}. 
A modified HMF is used \citep{Jokipii89}. 
The outer boundary is assumed to be at 120 AU.

The diffusion tensor is described in detail by \citet{Burger},
who used the same models for turbulence \citep*{Zank96},
but have adapted the coefficients to reflect some of the results from
the numerical simulations \citep*{Giacalone1,Giacalone2}.

The model used here gives latitudinal gradients in excellent agreement with
Ulysses observations, for both its value and its rigidity dependence
\citep{Burger}. The model also gives realistic radial dependence for
CR protons during consecutive solar minima, with the radial gradients
distinctly smaller in the $A>0$ than in the $A<0$ cycles. Detailed
fits to Pioneer and Voyager observations require that different
diffusion coefficients must be used for consecutive solar minimum
periods \citep{Potgieter00}.

\placefigure{fig:tilt}

The maximum extent of the HCS (tilt angle) has been used in modulation
models as a proxy for solar activity since the 1970's, when it was
realized that drifts were important. Figure \ref{fig:tilt} shows the
tilt angle\footnote{See URL
\url{http://quake.stanford.edu/$\sim$wso/}}, which is used as input in
our modulation calculations, for two of Hoeksema's models (classic
``L-model'' and a newer ``R-model'') vs.\ time. The upper and
lower curves represent his old and new model correspondingly
\citep[e.g.,][]{Hoeksema92,Hoeksema95}. The difference between the
models illustrates the error in the tilt angles one has to consider
when used in any modulation modeling.

\centerline{\psfig{file=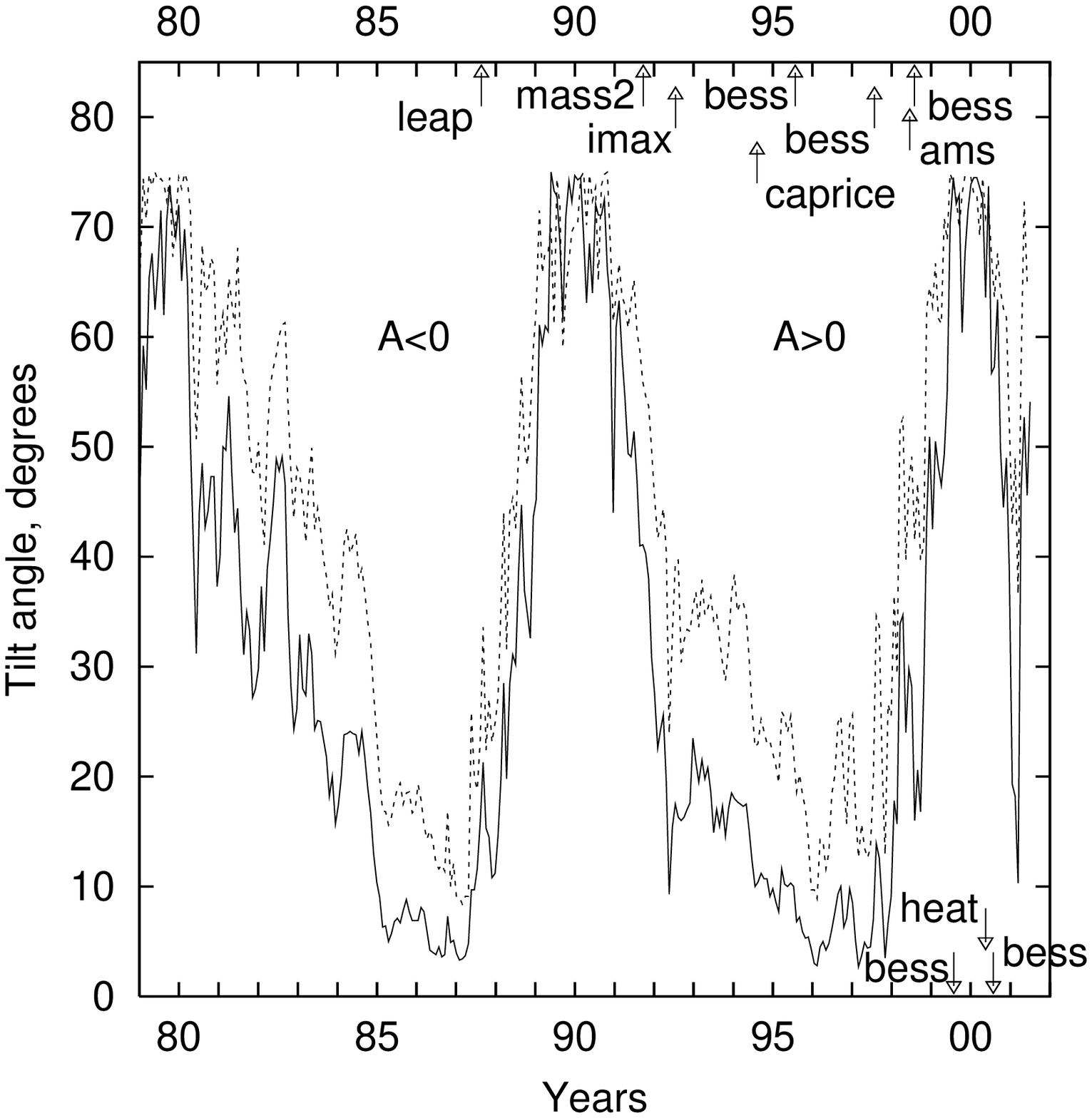,width=\twofigs,clip=}}
\figcaption[f2.ps]{ Heliospheric current sheet tilt angles for two
Hoeksema models \citep[e.g.,][]{Hoeksema92,Hoeksema95}, the classic
``L-model'' (dashes) and a newer ``R-model'' (solid).
\label{fig:tilt} \bigskip}

\section{Antiproton measurements} \label{sec:measurements}

\citet{Golden79} 
and independently \citet{bogomolov1} reported the first measurements of 
cosmic ray antiprotons. Both experiments 
employed a superconducting magnet for measuring
momentum and a gas Cherenkov detector to separate the background of
other negatively charged particles from antiprotons. 
These were the first attempts
to measure the flux near the 2 GeV peak expected
\citep{Gaisser74} in the differential spectrum from secondary cosmic
rays. The flux found was a factor of $\sim3$ above that expected and
caused a stir at the time. However, the statistics and systematic
errors in both the measurement and the theory of that time were such
that the excess was significant at about the $3\sigma$ level. With
the clarity of hindsight (see below), the \citet{Golden79}
and \citet{bogomolov1} measurements were
probably plagued by a background of negatively
charged particles that went undetected by the Cherenkov detector.

An annihilation technique \citep*{Buffington81} was used to make the
first alleged measurement of lower energy antiprotons ($< 1$ GeV).
This technique was designed to stop the antiprotons in a visualization
device. Antiprotons were identified by limiting their intrinsic
kinetic energy ($< 500$ MeV) and finding their rest energy ($\sim2$
GeV) from the number of annihilation prongs and large energy release.
We now know that this experiment must also have been plagued by a high
background of interacting particles because the flux of $\sim200$ MeV
antiprotons found exceeds that of more recent measurements by more
than an order of magnitude.

However, these papers stimulated the realization that antiprotons
could be produced by exotic processes such as the annihilation of
primordial black holes or dark matter in the galactic halo
(\citealp[e.g.,][]{carr,maki}; 
\citealp*{pbar_1,pbar_2,pbar_3,pbar_4,pbar_6,pbar_5,pbar_7,pbar_8}).

During the past 5 years, the BESS (Balloon-borne Experiments with a
Super-conducting Solenoid) Collaboration \citep{BESS_instrument} has
made a set of accurate new measurements that motivated this attempt
to make a more accurate model calculation. This impressive payload
now has data that extend from about 200 MeV (at the top of the
atmosphere) to 4 GeV. Importantly, this covers the energy range of
the expected peak in the antiproton distribution at 2 GeV where the
flux is highest. The first reported mass resolved measurements
\citep{old_pbar1,old_pbar2,Yoshimura95,mitchell} of antiprotons were
made at lower energies where determining both the momentum per unit
charge (rigidity) and velocity was easier. As can be seen from
equation (\ref{eq.50}) both are needed to determine the mass:
\begin{equation}
\label{eq.50}
m^2 \propto \rho^2\left(\beta^{-2}-1\right),
\end{equation}
where $\beta=v/c$, the velocity, and $\eta$, the curvature, are the
measured parameters. Rigidity, $\rho=1/\eta$, is inversely
proportional to the curvature. Uncertainties are given by
\begin{equation}
\label{eq.51}
\frac{\left(\Delta m\right)^2}{m^2} =
\frac{\left(\Delta\rho\right)^2}{\rho^2}+
\frac{\left(\Delta\beta\right)^2}{\beta^2\left(1-\beta^2\right)^2}.
\end{equation}
The statistical precision in these first measurements was poor
\citep{Yoshimura01}.

In BESS the charge sign and rigidity are determined by track
measuring drift chambers located in the uniform magnetic field inside
the bore of their superconducting magnet. Extending to higher energy
has required the use of not only the most precise time of flight
measurement but also the addition of aerogel Cherenkov detectors for
measuring velocity. The BESS team, through a series of annual
balloon campaigns in Lynn Lake, Manitoba, Canada, has continuously
improved the time resolution and in 1997 they added an aerogel
counter, extending the energy coverage beyond 2 GeV. In addition,
the data handling capacity has continuously improved to decrease dead
time and increase precision.

In this paper we use mainly BESS data collected in 1995 and 1997
\citep{Orito00}. However in one plot (\S~\ref{sec:variations},
Fig.\ \ref{fig:pbar_combined}), we have combined the data from those
BESS flights with their 1998 data \citep{Maeno01} after correcting it
for increased solar modulation level. The flights in 1995 and 1997
took place near solar minimum conditions while the 1998 flight was
just after solar modulation minimum (Fig.\ \ref{fig:tilt}). 
We compare our predictions for different heliospheric modulation levels with
data from BESS flights in 1995-97, 1998, 1999, and 2000 \citep{Orito00,Maeno01,asaoka01}
in Figures \ref{fig:prot_cycle}-\ref{fig:pbar2p}. The
BESS flight in year 2000 provides the first accurate measurement of 
the $\bar p/p$ ratio for the negative polarity solar cycle.

The statistical precision with which the
flux at 2 GeV is known is now better than 10\% ($1\sigma$). The data
are now good enough to detect solar modulation effects at this energy
at about the same magnitude.
The BESS group has carried out an extensive calibration of their
instrument to check for ways to reduce sources of systematic error.
They report a systematic uncertainty in the antiproton flux measurements to
be about 5\% \citep{bess_calibration}.

Data at energies above 4 GeV on the steeply falling part of the
antiproton spectrum are more difficult to obtain. We have used the
results reported by the MASS group \citep{hof,basini} from a flight
in 1991 based on the original \citet{Golden79} payload but with an
improved gas Cherenkov detector; we have used their most recent
analysis \citep{Stoc01}. These data have error bars that extend up
and down by a factor of $\sim2$. We can say that they are consistent
with antiprotons being of secondary origin but are not precise enough
to place constraints on the model presented here.

\section{New calculations} \label{sec:calc}

\subsection{Local proton and helium spectra} \label{sec:spectra}

Secondary antiprotons are produced in collisions of energetic protons
and helium nuclei with interstellar gas, so the interstellar spectra
of these nuclei is fundamental to calculating antiproton spectra
expected at Earth. A major problem in determination of the LIS CR
spectrum is the effect of heliospheric modulation that has been
discussed in detail in \S~\ref{sec:modulation}. Inverting equation
(\ref{eq.40}) is \emph{not} well-defined making the accurate
derivation of the LIS spectrum a complicated task. However at energies
above the antiproton production threshold, at 10--30 GeV/nucleon, the
heliospheric modulation is weak. We thus try to get an approximate
LIS spectrum using the force-field approximation \citep{forcefield}.
The appropriate modulation potential ($\Phi$) has been chosen using
CLIMAX neutron monitor data \citep{Badhwar96}.

Spectra given in the literature are quoted in different ways,
sometimes as power laws in kinetic energy, sometimes rigidity or total
energy. Also different measurements have different systematic
uncertainties in flux. In order to best determine the asymptotic slope
of the locally observed spectrum we have summarized all recent
measurements and fitted them assuming a LIS spectrum which is a
power-law in kinetic energy. (We note that fitting to a power-law in
rigidity yields the spectrum which is too steep to agree with high
energy data by Sokol and JACEE.)

To get an idea of which part of the nucleon spectrum contributes most
to antiproton production, we have made runs in which we cut the
nucleon spectrum at different energies. Our analysis shows that
$\sim97$\% of all antiprotons below 6 GeV are produced by nucleons
below $\sim200$ GeV/nucleon. The nucleons at $\leq20$ GeV/nucleon
yield about 1/3 of all antiprotons $\leq1$ GeV, nucleons of $\leq50$
GeV/nucleon yield about 80\% of all antiprotons $\leq2$ GeV, and 90\%
of antiprotons $\leq4$ GeV are produced by nucleons of $\leq100$
GeV/nucleon. Therefore the spectra at moderate energies, up to
$\sim200$ GeV/nucleon, are of most importance.

\placetable{table2}

\placetable{table3}


\begin{table*}[!th] 
\tablecolumns{8}
\tablewidth{0mm}
\tabletypesize{\footnotesize}

\caption{Parameters of the LIS proton spectrum\tablenotemark{a}~ as derived in present paper.
\label{table2}}
\begin{tabular}{lccccclc}\hline\hline\noalign{\smallskip}

\colhead{} & 
\colhead{Fitting interval} & 
\colhead{Normalization,} & 
\colhead{Power-law} &
\colhead{Modulation} & 
\colhead{Fit qua-} & 
\colhead{Flight date,} & 
\colhead{Data}
\\
\colhead{Instrument} & 
\colhead{$E_{kin}$, GeV} &
\colhead{(m$^2$ s sr GeV)$^{-1}$  } & 
\colhead{index} &
\colhead{potential, MV} & 
\colhead{lity, $\chi^2_n$} & 
\colhead{yymmdd} & 
\colhead{reference}
\startdata
LEAP          & 
20--100       &
---           & 
$2.69\pm0.04$ &  
550           & 
0.80          & 
870821        & 1
\smallskip\\
MASS2         &
20--100       &
$(1.93\pm0.15)\times10^4$ &
$2.82\pm0.03$ &
1200          &
0.53          &
910923        & 2
\smallskip\\
IMAX          &
20--200       &
$(1.08\pm0.15)\times10^4$ &
$2.66\pm0.04$ &
750           &
0.26          &
920716-17     & 3
\smallskip\\
CAPRICE       &
20--200       &
$(1.55\pm0.19)\times10^4$ &
$2.80\pm0.03$ &
600           &
2.54          &
940808-09     & 4
\smallskip\\
AMS           &
20--200       &
$(1.82\pm0.21)\times10^4$ &
$2.79\pm0.03$ &
550           &
0.13          &
9806          & 5
\smallskip\\
BESS          &
20--120       &
$(1.61\pm0.13)\times10^4$ &
$2.75\pm0.03$ &
550           &
0.05          &
980729-30     & 6
\medskip\\
\multicolumn{2}{l}{Weighted average}&
$(1.58\pm0.08)\times10^4$&
$2.76\pm0.01$ 
\\ 
\enddata
\tablenotetext{a}{Assuming power-law in kinetic energy LIS spectrum.}
\tablerefs{(1) \citet{seo91}; (2) \citet{mass2}; (3) \citet{Menn00}; (4) \citet{Boez99};\\
           (5) \citet{p_ams}; (6) \citet{Sanu00}.}

\end{table*}


\begin{table*}[!th]
\tablecolumns{8}
\tablewidth{0mm}
\tabletypesize{\footnotesize}

\caption{Parameters of the LIS helium spectrum\tablenotemark{a}~ as derived in present paper.
\label{table3}}
\begin{tabular}{lccccclc}\hline\hline\noalign{\smallskip}

\colhead{} & 
\colhead{Fitting interval} & 
\colhead{Normalization,} & 
\colhead{Power-law} &
\colhead{Modulation} & 
\colhead{Fit qua-} & 
\colhead{Flight date,} & 
\colhead{Data}
\\
\colhead{Instrument} & 
\colhead{$E_{kin}$, GeV/n} &
\colhead{(m$^2$ s sr GeV/n)$^{-1}$  } & 
\colhead{index} &
\colhead{potential, MV} & 
\colhead{lity, $\chi^2_n$} & 
\colhead{yymmdd} & 
\colhead{reference}
\startdata
LEAP          &
10--100       &
---           &
$2.69\pm0.09$ &
550           &
---           &
870821        & 1
\smallskip\\
MASS2         &
10--50        &
$686\pm130$   &
$2.79\pm0.07$ &
1200          &
0.44          &
910923        & 2
\smallskip\\
IMAX          &
10--125       &
$600\pm120$   &
$2.70\pm0.07$ &
750           &
0.69          &
920716-17     & 3
\smallskip\\
CAPRICE       &
10--100       &
$590\pm124$   &
$2.73\pm0.07$ &
600           &
0.66          &
940808-09     & 4
\smallskip\\
AMS           &
10--100       &
$653\pm56$    &
$2.72\pm0.03$ &
550           &
0.39          &
9806          & 5
\smallskip\\
BESS          &
10--50        &
$640\pm139$   &
$2.67\pm0.07$ &
550           &
0.14          &
980729-30     & 6
\medskip\\
\multicolumn{2}{l}{Weighted average}&
$641\pm53$    &
$2.72\pm0.01$ 
\\ 
\enddata
\tablenotetext{a}{Assuming power-law in kinetic energy per nucleon LIS spectrum.}
\tablerefs{(1) \citet{seo91}; (2) \citet{mass2}; (3) \citet{Menn00}; (4) \citet{Boez99};
           (5) \citet{he_ams}; (6) \citet{Sanu00}.}

\end{table*}

Tables \ref{table2} and \ref{table3} show our fits to recent
measurements of CR protons and helium. (The power-law in kinetic energy
has been modulated with appropriate modulation potential and then
fitted to the data.) 
From the parameters fitted to each individual set of measurements
we calculate the weighted averages. The reduced $\chi^2_n$ ($\chi^2$
per the degree of freedom) shows the quality of the fits.

The fitted parameters are not unique since the modulation potential is
not well determined, and on account of the systematic and statistical
errors. We therefore use also representative high energy data, above
$\sim100$ GeV/nucleon, where the modulation has no effect, for a
cross check. In particular, our composite proton spectrum,
$1.60\times10^4 E_{kin}^{-2.75}$ m$^{-2}$ s$^{-1}$ sr$^{-1}$
GeV$^{-1}$ (Fig.\ \ref{fig:protons}, bold dashes), passes through Sokol
\citep{Sokol} and JACEE data points \citep{jacee}. The index agrees
well with that determined by \citet*{ryan72}, $2.75\pm0.03$, and HEGRA
below the knee \citep{hegra}, $2.72_{-0.03}^{+0.02}\pm0.07$. The
derived LIS proton spectrum thus agrees very well with all the data in
the range $\sim10-10^5$ GeV.

Our calculation of the nucleon spectrum is based on the solution of
the diffusion equation (\ref{eq.1}), which gives the spectrum in every
cell of the spatial grid. Shock acceleration models predict that the
injection CR spectrum can be approximated by a power-law in rigidity
which may steepen at low energies  (e.g., Ellison, Slane, \& Gaensler
2001). Hence we adopt a power-law in rigidity, if necessary with a
break,  for the injection spectrum. The local measurements are
conventionally used for comparison and normalization of the propagated
spectrum. While there are some indications that the local interstellar
(LIS) nucleon spectrum may be not representative \citep{SMR00}, at
high energies where the diffusion is fast and energy losses become
small it should be close to that observed near the Earth. This is also
indicated by studies of Galactic diffuse emission, antiprotons, and
positrons \citep{hunter97,MSR98,SMR00}. While our propagation
calculations yield LIS spectra which can not be described by a single
power-law, the estimates together with data will serve as useful
guidelines.

The propagated helium spectrum with an appropriate normalization
matches after modulation the data better than the averaged LIS
spectrum from Table \ref{table3}. The latter is somewhat too low when
compared with high-energy data by Sokol and JACEE.


\begin{table*}[!t]
\tablecolumns{7}
\tablewidth{0mm}
\tabletypesize{\footnotesize}

\caption{Propagation parameter sets.\label{table4}}
\begin{tabular}{lcccccc}\hline\hline\noalign{\smallskip}

\colhead{} & 
\colhead{} & 
\multicolumn{2}{c}{Diffusion coefficient\tablenotemark{a}} & 
\colhead{} &
\multicolumn{2}{c}{Reacceleration\tablenotemark{b}\,/Convection}
\\
\cline{3-4}\cline{6-7}
\colhead{Model} & 
\colhead{Injection index, $\gamma$\tablenotemark{c}} &
\colhead{$D_0$, cm$^2$ s$^{-1}$} & 
\colhead{Index, $\delta$} &
\colhead{} &
\colhead{$v_A/\sqrt w$, km s$^{-1}$} & 
\colhead{$dV/dz$, km s$^{-1}$ kpc$^{-1}$}
\startdata
Diffusive            \\
Reacceleration (DR)  &
2.43                 &
$6.10\times10^{28}$  & 
0.33                 &
 & 
30                   &
---                  
\smallskip\\

Diffusive            \\
Reacceleration       \\
with Break (DRB)     &
1.93/2.43\tablenotemark{d}&
$6.10\times10^{28}$  & 
0.33                 &
 & 
30                   &
---                  
\smallskip\\

Minimal Reacceleration\\
plus Convection (MRC)&
2.43                 &
$4.30\times10^{28}$  & 
0.33                 & 
 &
17                   &
10                   
\smallskip\\

Plain Diffusion (PD) &
2.16                 &
$3.10\times10^{28}$  & 
0.60                 &
 &
---                  &
---                  
\smallskip\\


Diffusion plus       \\
Convection (DC)      &
2.46/2.16\tablenotemark{e}&
$2.50\times10^{28}$  & 
0/0.60\tablenotemark{a}&
 &
---                  &
10                   
\\ 
\enddata
\tablecomments{Adopted halo size $z_h=4$ kpc.}
\tablenotetext{a}{$\rho_0=4$ GV, 
   index $\delta$ is shown below/above $\rho_0$.}
\tablenotetext{b}{$v_A$ is the Alfv\'en speed, and
$w$ is defined as the ratio of MHD wave energy density to magnetic field energy density.}
\tablenotetext{c}{For a power-law in rigidity, $\propto\rho^{-\gamma}$.}
\tablenotetext{d}{Index below/above rigidity 10 GV.}
\tablenotetext{e}{Index below/above rigidity 20 GV.}

\vspace{-2\baselineskip}
\end{table*}

\subsection{Propagation models and parameters} \label{sec:prop}

To investigate the range of interstellar spectra and propagation
parameters we have run a large number of models. Here we consider five 
cases (see Table \ref{table4}), which differ in galactic propagation. 
The diffusive reacceleration (DR) model has been chosen since
it has been very successful in the description of the propagation of CR nuclei; it 
reproduces the sharp peak in the secondary to primary nuclei ratios in a physically
understandable way without breaks in the diffusion coefficient and/or
the injection spectrum. We consider also a 
diffusive reacceleration model with break in the injection
spectrum (DRB) in order to match the LIS proton and He spectra.
The minimal reacceleration plus convection (MRC) model uses
reduced reacceleration, and in addition some convection is invoked to
reproduce the B/C ratio. We compute also a plain diffusion model (PD)
for reference, although it does not fit B/C.
In the fifth model we use diffusion and convection (DC) 
with breaks in the diffusion coefficient and the injection index in
order to construct a model capable of describing all the CR (particle)
data simultaneously.

\placetable{table4}

In all models,
the injection spectrum was chosen to reproduce the local CR
measurements (see \S~\ref{sec:spectra}). The source abundances of all
isotopes $Z\leq28$ are given in \citet{SM01}.
The propagation parameters have been fixed using the B/C ratio.
Because of the cross section fits to the main Be and B production
channels and renormalization to the data where it exists, the
propagation parameters show only weak dependence ($\sim10$\% change
in the diffusion coefficient) on the cross section parametrization
\citep{ST-code,W-code}. We thus use \citeauthor{W-code} cross
section code in our calculations.

The halo size has been set to $z_h=4$ kpc, which is
within the range $z_h=3-7$ kpc derived using the GALRPOP code and the
combined measurements of radioactive isotope abundances, $^{10}$Be,
$^{26}$Al, $^{36}$Cl, and $^{54}$Mn \citep{SM01}. This is also in
agreement with our newer estimate $z_h=4-6$ kpc \citep*{MMS01}.
Note that the exact value of $z_h$ is unimportant
for antiproton calculations provided that the propagation parameters
are tuned to match the B/C ratio.

\begin{figure*}[!thb] 
\psfig{file=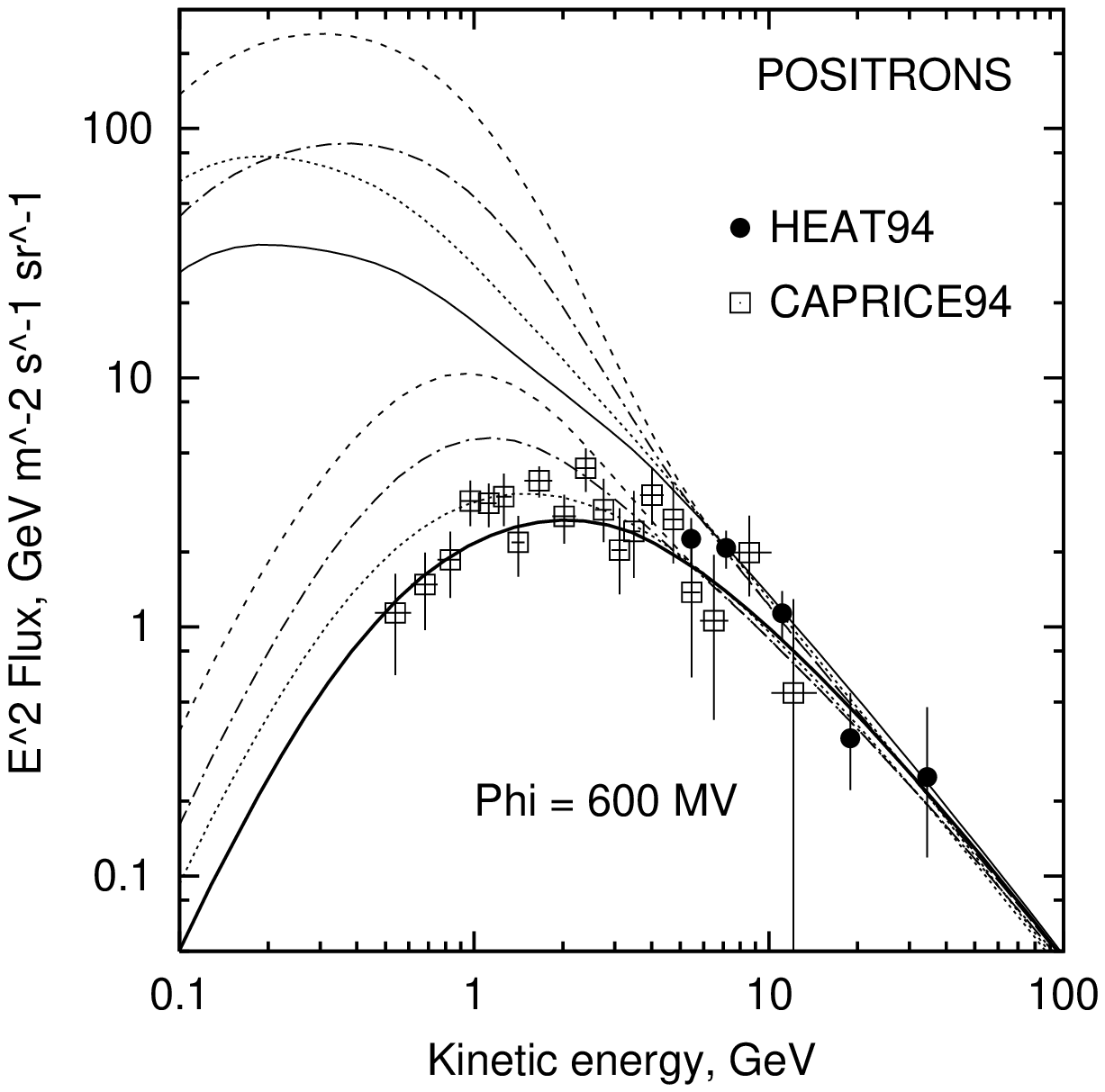,width=\twofigs,clip=}
\psfig{file=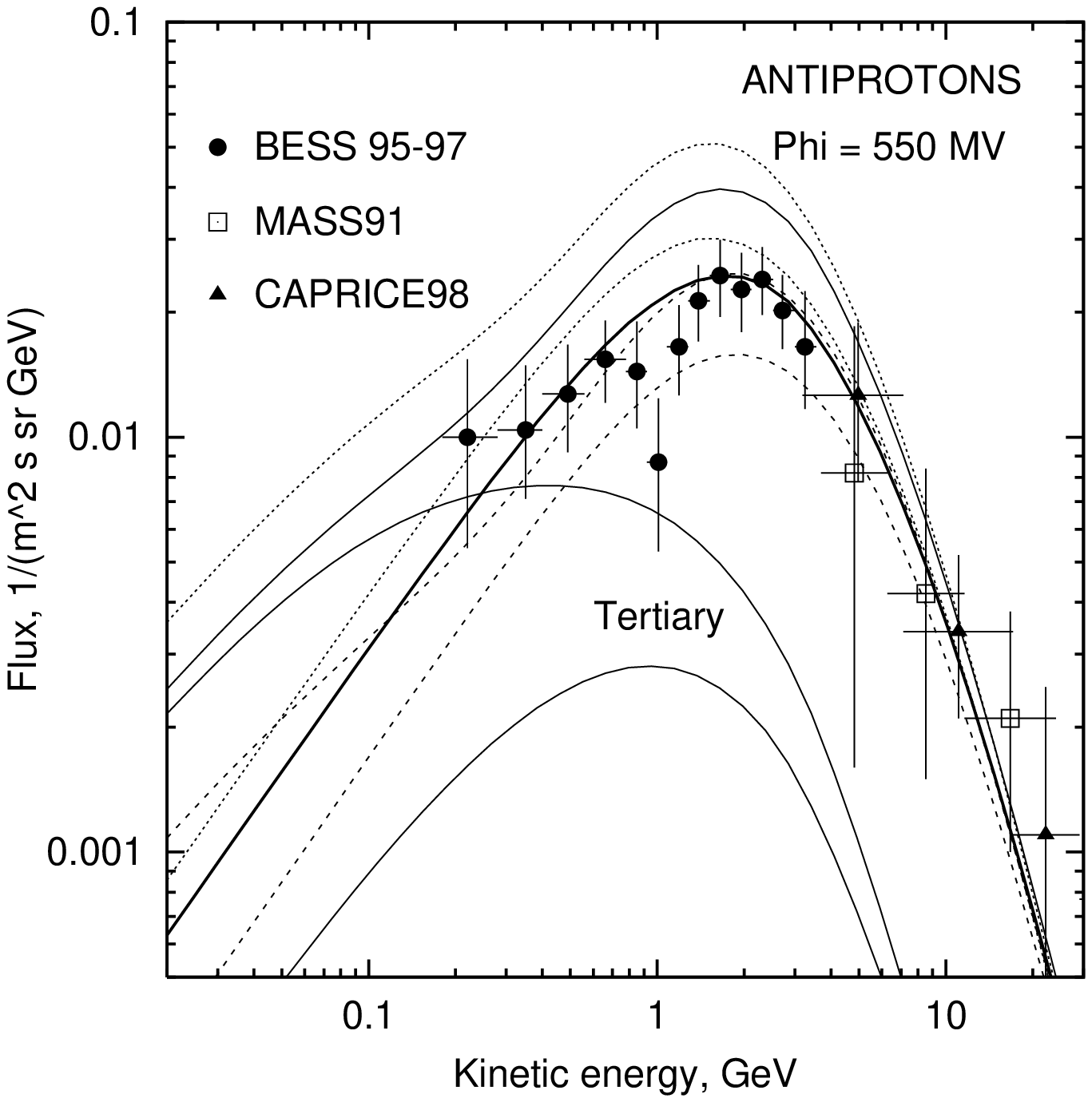,width=\twofigs,clip=}
\begin{minipage}[tl]{\wcap}
\figcaption[f6.ps]{ Calculated positron flux: upper curves LIS, lower
modulated (force field, $\Phi=600$ MV). The lines are coded as in
Figure \ref{fig:BC}. Positron flux as calculated in DRB model is
shown by dash-dots. Data: HEAT94 \citep{barwick98}, CAPRICE94
\citep{boezio00}.
\label{fig:pos} \bigskip}
\end{minipage} \hfill
%
\begin{minipage}[tr]{\wcap}
\figcaption[f7.ps]{ Calculated antiproton spectrum, upper curves -- LIS,
modulation was made with $\Phi=550$ MV (force field, lower curves).
The lines are coded as in Figure \ref{fig:BC}. Data: BESS
\citep{Orito00}, MASS91 \citep{Stoc01}, CAPRICE98 \citep{boezio01}.
The top curves are the total secondary $\bar p$'s. The two lowest
curves marked ``tertiary'' show separately the LIS spectrum and
modulated ``tertiary'' component in DC model.
\label{fig:pbars} \bigskip}
\end{minipage}
\end{figure*}

Our results are plotted in Figures \ref{fig:BC}--\ref{fig:pbars}. The
upper curve is always the LIS spectrum, except in the B/C ratio plot (Fig.\
\ref{fig:BC}) where the \emph{lower} curves are the LIS ratio. The
modulation on these plots has been done using the force-field
approximation. For the B/C ratio (Fig.\ \ref{fig:BC}) in the DC
model the result of modulation in the drift model is also shown
though the difference (force-field approximation vs.\ drift model)
is small. 

\placefigure{fig:BC}

\begin{figure*}[!p]
\centerline{\psfig{file=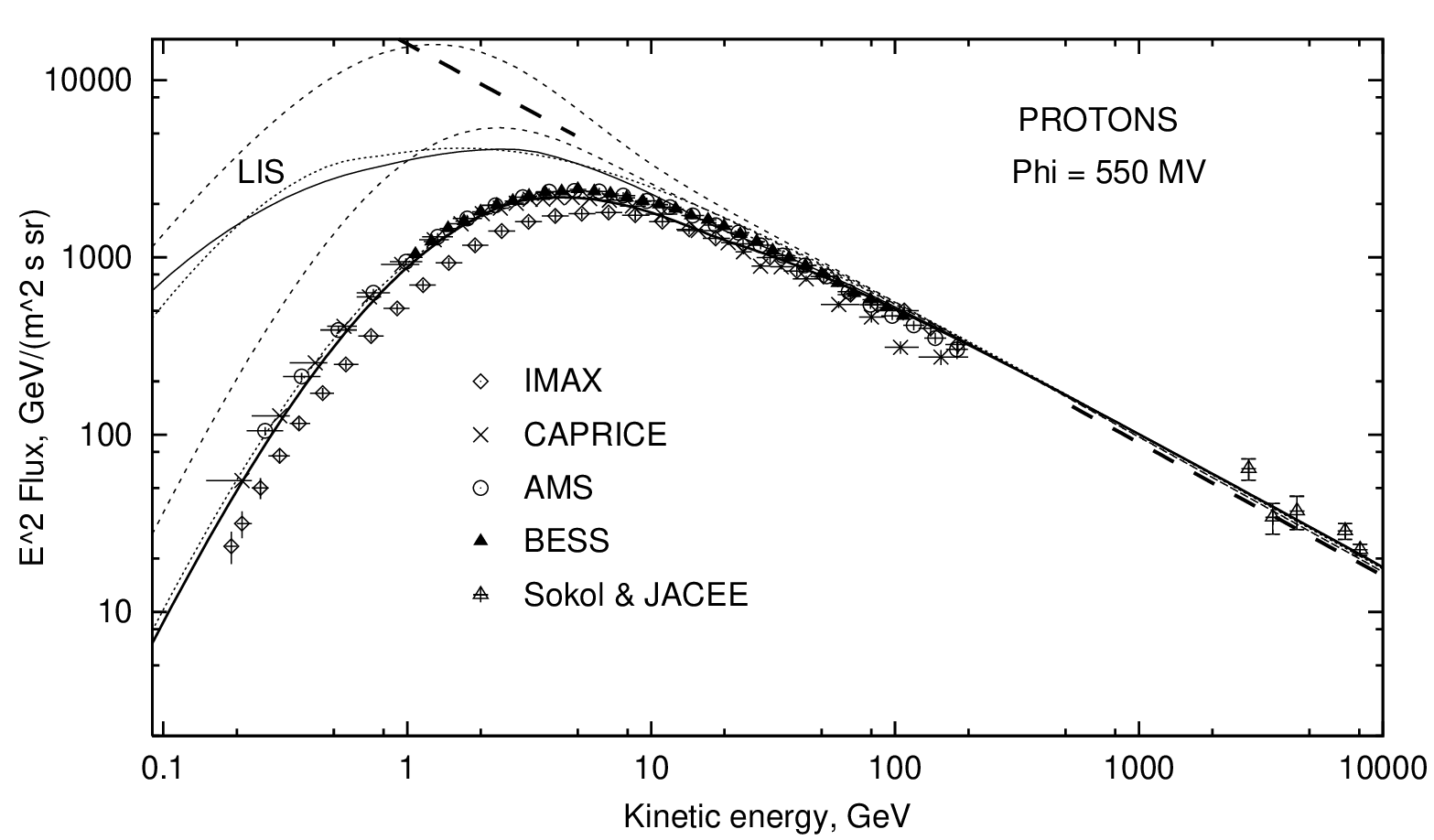,width=\onefig,clip=}\smallskip}
\figcaption[f4.ps]{ Calculated proton interstellar spectrum (LIS) and
modulated spectrum (force field, $\Phi=550$ MV). The lines are coded
as in Figure \ref{fig:BC}. Bold dashed line shows our composite
spectrum as derived from measurements $1.60\times10^4 E_{kin}^{-2.75}$
m$^{-2}$ s$^{-1}$ sr$^{-1}$ GeV$^{-1}$ (not shown between 5--500 GeV
for clarity).  Data: IMAX \citep{Menn00}, CAPRICE \citep{Boez99}, AMS
\citep{p_ams}, BESS \citep{Sanu00}, Sokol \citep{Sokol}, and JACEE
\citep{jacee}.
\label{fig:protons} \bigskip}
%
\centerline{\psfig{file=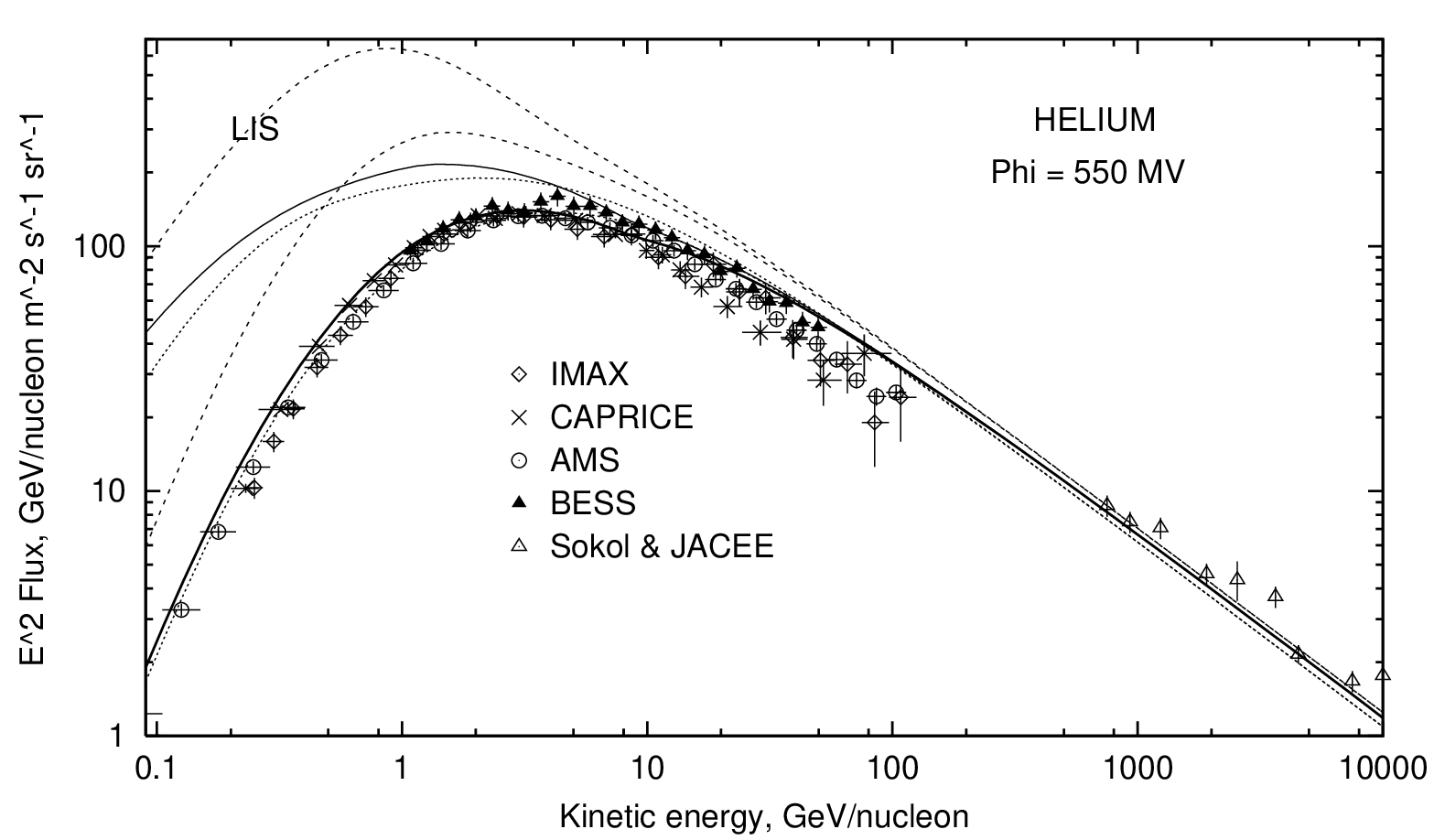,width=\onefig,clip=}\smallskip}
\figcaption[f5.ps]{ Calculated He interstellar spectrum (LIS) and
modulated spectrum (force field, $\Phi=550$ MV). The lines are coded
as in Figure \ref{fig:BC}. Data: IMAX \citep{Menn00}, CAPRICE
\citep{Boez99}, AMS \citep{he_ams}, BESS \citep{Sanu00}, Sokol
\citep{Sokol}, and JACEE \citep{jacee}.
\label{fig:He} \bigskip}
\end{figure*}

\placefigure{fig:protons}

\placefigure{fig:He}

\placefigure{fig:pos}

\placefigure{fig:pbars}

The ``tertiary'' antiprotons (inelastically scattered secondaries),
significant at the lowest energies, are important in interstellar
space, but make no difference when compared with measurements in the
heliosphere (Fig.\ \ref{fig:pbars}).

The model with the diffusive reacceleration (DR) reproduces the sharp
peak in secondary to primary nuclei ratios \citep{SM98,SM01}.
However, this model produces
a bump in proton and He spectra at $\sim2$ GeV/nucleon that is not
observed.\footnote{A similar bump is produced also in the electron
spectrum at 1 GeV.} This bump can be removed by choosing an injection
spectrum that is flatter at low energies \citep[see,
e.g.,][]{jones01}. There are however some problems with secondaries
such as positrons and antiprotons which are difficult to manage. A
similar bump appears in the positron spectrum at $\sim1$ GeV, and the
model underproduces antiprotons at 2 GeV by more than 30\%.

In order to fit the proton/Helium spectra in a diffusive 
reacceleration model we need to introduce a break in the injection
spectrum, with a smaller index below 10 GV (model DRB, \emph{not} 
shown in the plot for clarity).
This break in the injection spectrum of primaries 
results in smaller bump in the secondary posiron spectrum 
(dash-dots in Fig.\ \ref{fig:pos}),
though the predicted spectrum still shows a significant excess of 
a factor at least two over the measurements. However, it produces no effect 
on secondary antiprotons because of their higher production 
threshold energy. Taken
together these cases provide evidence against ``strong''
reacceleration\footnote{ We define that the reacceleration is
``strong'' if the model is able to match the B/C ratio without
invoking other mechanisms such as convection and/or breaks in the
diffusion coefficient.} in the ISM.

\centerline{\psfig{file=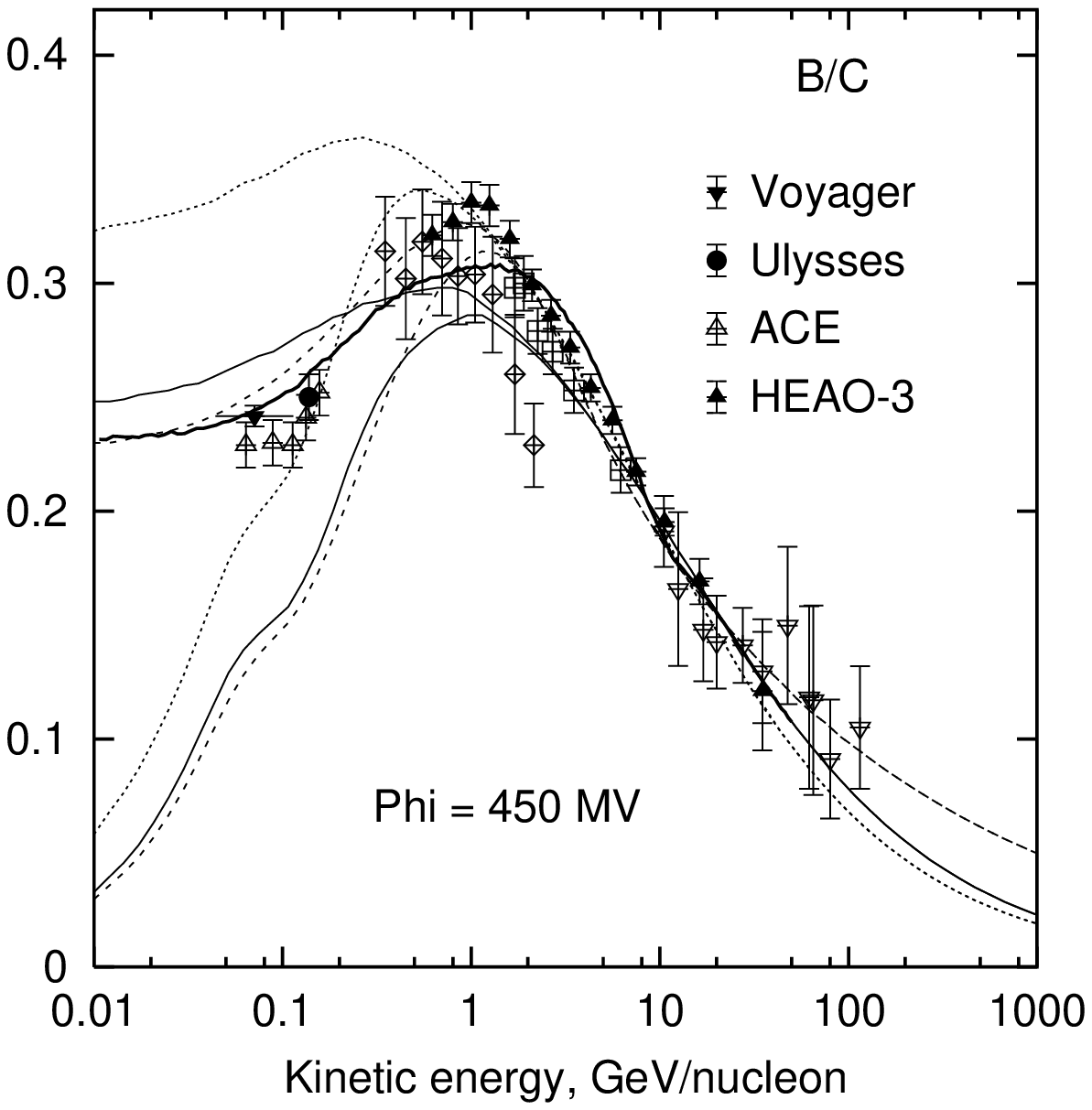,width=\twofigs,clip=}}
\figcaption[f3.ps]{ B/C ratio as calculated in DC, DR, and PD models for
$z_h=4$ kpc. Lower curves -- interstellar, upper -- modulated
($\Phi=450$ MV). The line coding is: solid lines -- DC model, dashes
-- DR, dots -- PD. The thick solid line (DC model) corresponds to the
drift model calculations of heliospheric modulation for $A>0$ and tilt
angle $5^\circ$. Data below 200 MeV/nucleon: ACE \citep{ace}, Ulysses
\citep*{ulysses_bc}, Voyager \citep*{voyager}; high energy data:
HEAO-3 \citep{Engelmann90}, for other references see
\citet{StephensStreitmatter98}.
\label{fig:BC}} 

Another model combining reduced reacceleration and convection (MRC)
also produces too few antiprotons.

Using a plain diffusion model (PD) we can get good agreement with B/C
above few GeV/nucleon, with nucleon spectra and positrons, but this
model overproduces antiprotons at 2 GeV by $\sim20$\% and contradicts
the secondary/primary nuclei ratio (B/C) below 1 GeV/nucleon.

A diffusion model with convection (DC) is our ``best fitting model''.
It reproduces all the particle data ``on average'', although it has
still some problem with reproduction of the sharp peak in the B/C
ratio. In this model a flattening of the diffusion coefficient
($\delta=0$) below 4 GV is required to match the B/C ratio at low
energies. A possible physical origin for the behaviour of the diffusion
coefficient is discussed in \S~\ref{sec:discussion}.

To better match primaries ($p$, He) in the DC model, we introduced a
steeper injection spectrum below 20 GV to compensate the break in
the diffusion coefficient, which tends to flatten the spectrum at low energies.
Such a steepening in the injection spectrum, however, has almost no effect on
secondaries ($\bar p$, $e^+$).
The existence of a sharp
upturn below few GeV/nucleon is in fact predicted from SNR shock
acceleration theory \citep*{ellison}; this is a transition region
between thermal and non-thermal particle populations in the shock. Our
model does not require a large break (0.3 in index is enough, see
Table \ref{table4}). More discussion is given in \S\
\ref{sec:discussion} and \S~\ref{sec:concl}.

The drift model has been used for heliospheric modulation of all
spectra obtained in DC model ($p$, He, B/C, $e^+$, $\bar p$). The same
modulation parameter set has been applied to all the species and the
agreement is good. The B/C ratio (Fig.\ \ref{fig:BC}) is virtually
insensitive to the modulation level, and it remains the same for a wide
variety of tilt angles.

\subsection{Discussion} \label{sec:discussion}

Our ``best-fit'' model (DC) with a constant diffusion coefficient
below 4 GV suggests some change in the propagation mode. Propagation
and scattering of the high energy particles on magnetic turbulence is
described by diffusion with the diffusion coefficient increasing
with energy. The growth of the diffusion coefficient depends on the
adopted spectrum of magnetic turbulence and is typically in the range
$0.3-0.6$. However, at low energies particles could propagate
following the magnetic field lines rather than scatter on magnetic
turbulence. This change in the propagation mode may affect the
diffusion coefficient making it less dependent on energy. Since the
magnetic field lines are essentially tangled such a process can still
behave like diffusion \citep{berezinskii}.

One more unknown variable is the CR spectrum in the distant regions of
the Galaxy. The LIS nucleon spectrum is studied quite well by direct
measurements at high energies where solar modulation effects are
minimal. Meanwhile the ambient CR proton spectrum on the large scale
remains unknown. The most direct test is provided by diffuse \grays.
However there is the well known puzzle of the GeV excess in the EGRET
diffuse \gray\ spectrum \citep{hunter97} which makes a direct
interpretation in terms of protons problematic, and possibly
inverse-Compton emission from a hard electron spectrum is responsible
\citep{SMR00}; for this reason we do not consider \grays\ further here.

An interesting independent possibility to test the interstellar nucleon 
spectrum at different energies is provided by antiproton and positron data
taken together. While the spectra of antiprotons and positrons at above
$\sim1$ GeV allow conclusions on the nucleon spectrum above $\sim10$ GeV
\citep{MSR98}, the positron spectrum below 1 GeV is sensitive to the
nucleon spectrum below the antiproton production threshold $\sim10$ GeV.
The problem here is that the propagation parameters are not known
accurately enough and this affects the predicted spectrum of positrons at low 
energies.

\medskip
\section{Variations of proton and antiproton spectra over the solar
cycles} \label{sec:variations}

\begin{figure*}[!p]
\centerline{\psfig{file=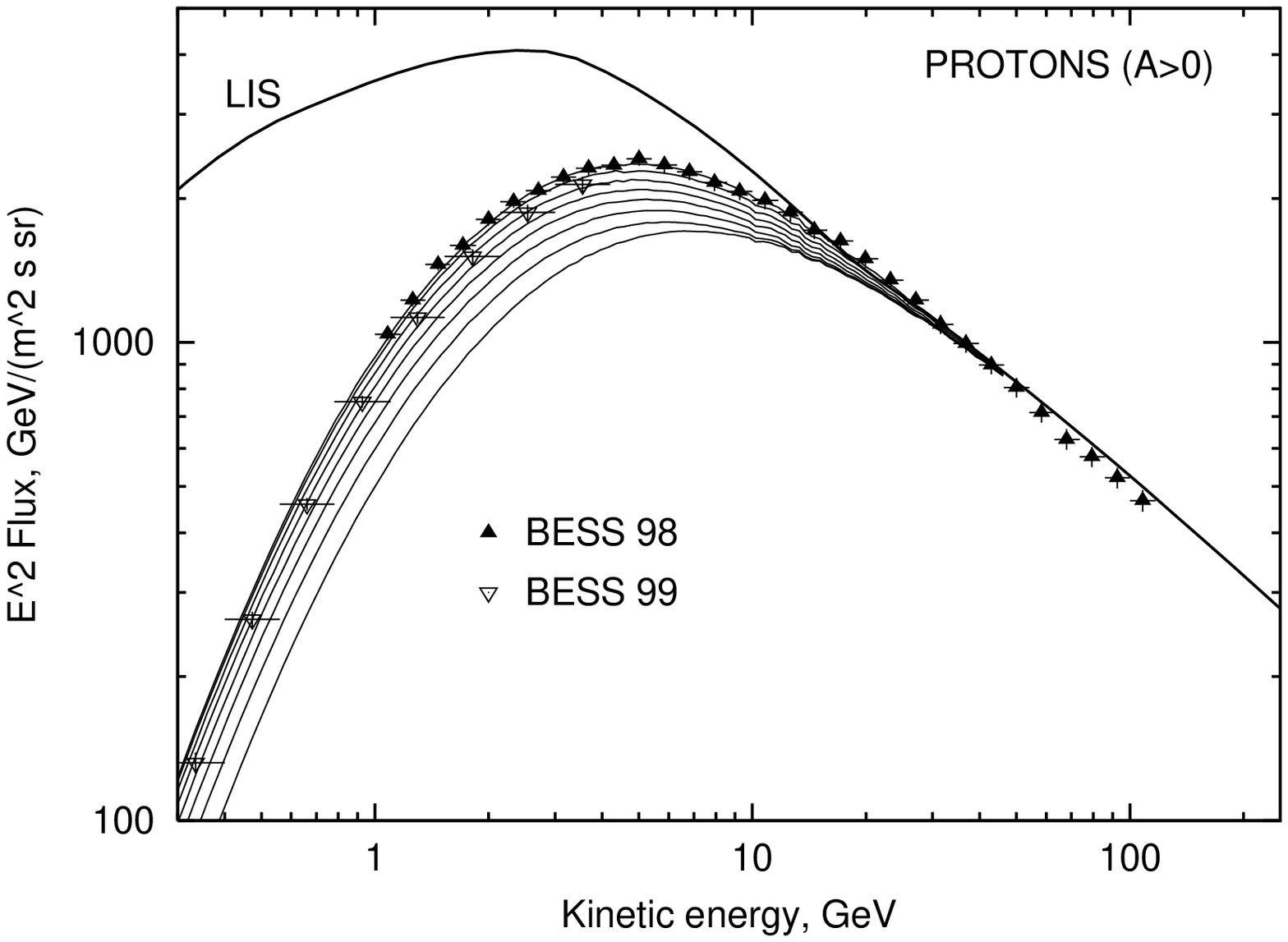,width=\onefig,clip=}}
\centerline{\psfig{file=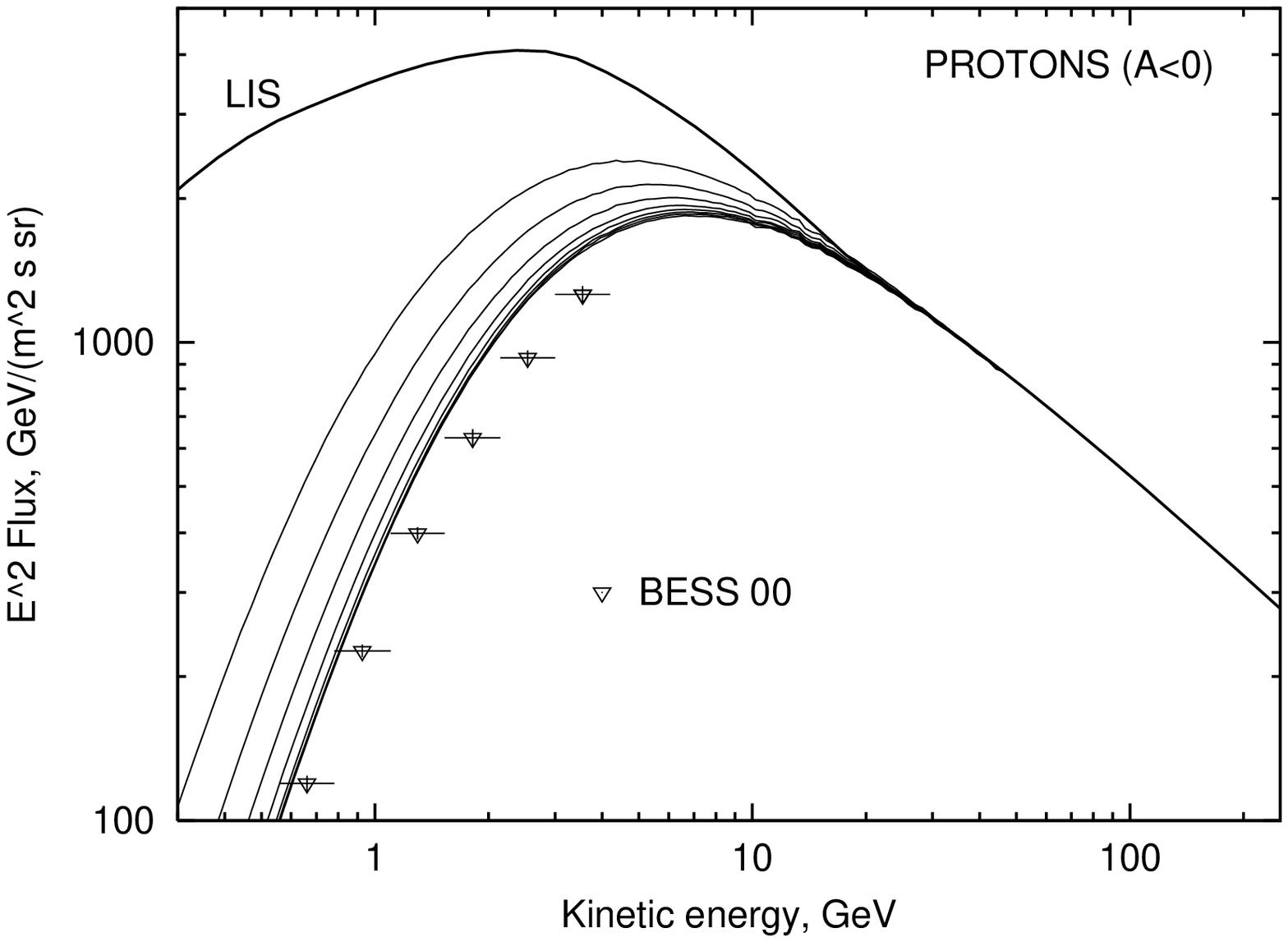,width=\onefig,clip=}}
\figcaption[f8a.ps,f8b.ps]{ Calculated proton LIS and modulated spectra
for the two magnetic polarity dependent modulation epochs, $A>0$
(top) and $A<0$ (bottom). Tilt angle from top to bottom: $5^\circ,
15^\circ, 25^\circ, 35^\circ, 45^\circ, 55^\circ, 65^\circ, 75^\circ$.
The tilt angle corresponding to BESS-98 data is
$\sim\!5^\circ\!-15^\circ$ ($A>0$) depending on the coronal field
model.
For discussion of BESS-99 and BESS-00 data see text.
Data: BESS-98 \citep{Sanu00}, BESS-99, BESS-00 \citep{asaoka01}.
\label{fig:prot_cycle} \bigskip}
\end{figure*}

\begin{figure*}[!p]
\centerline{\psfig{file=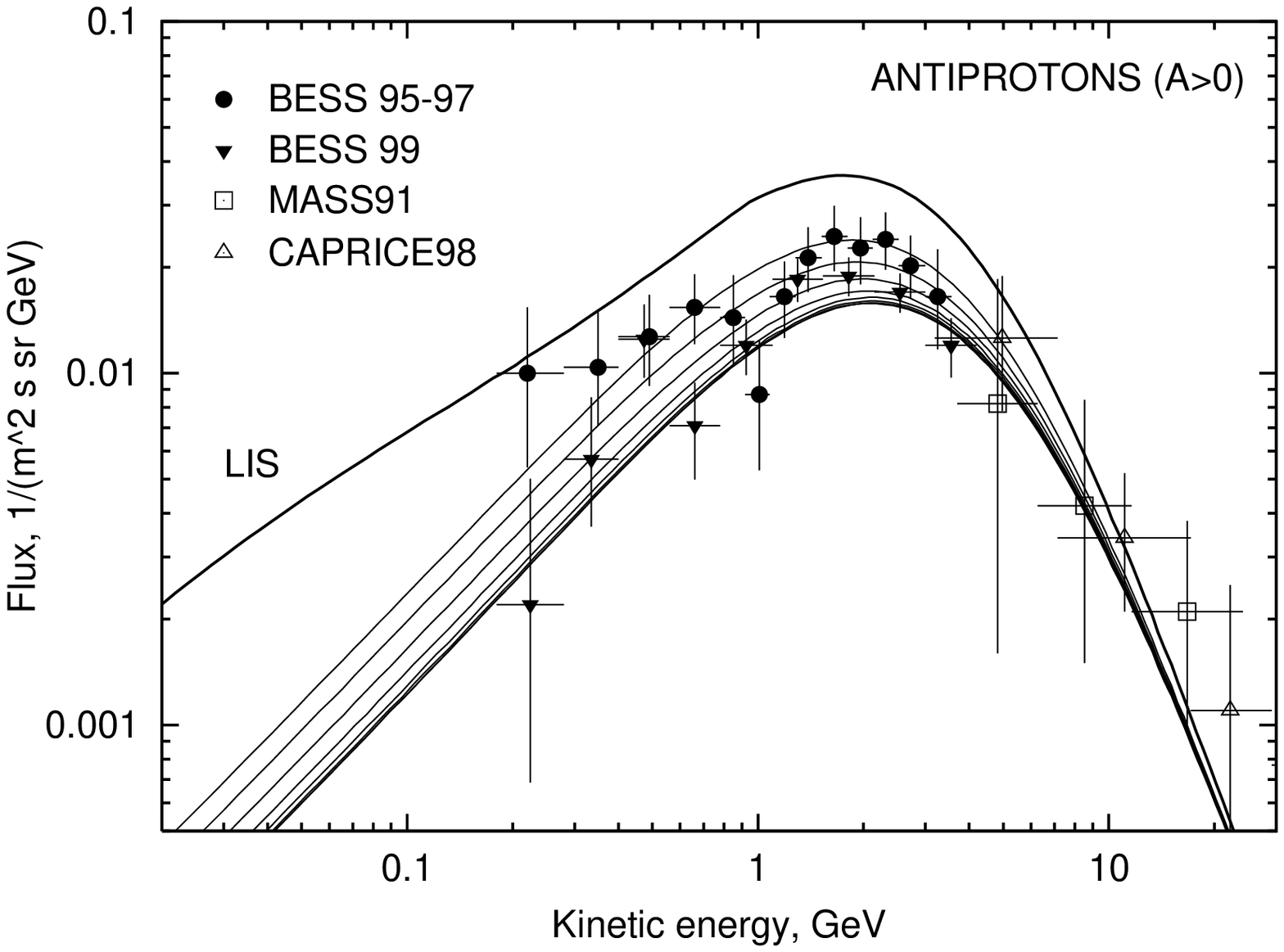,width=\onefig,clip=}}
\centerline{\psfig{file=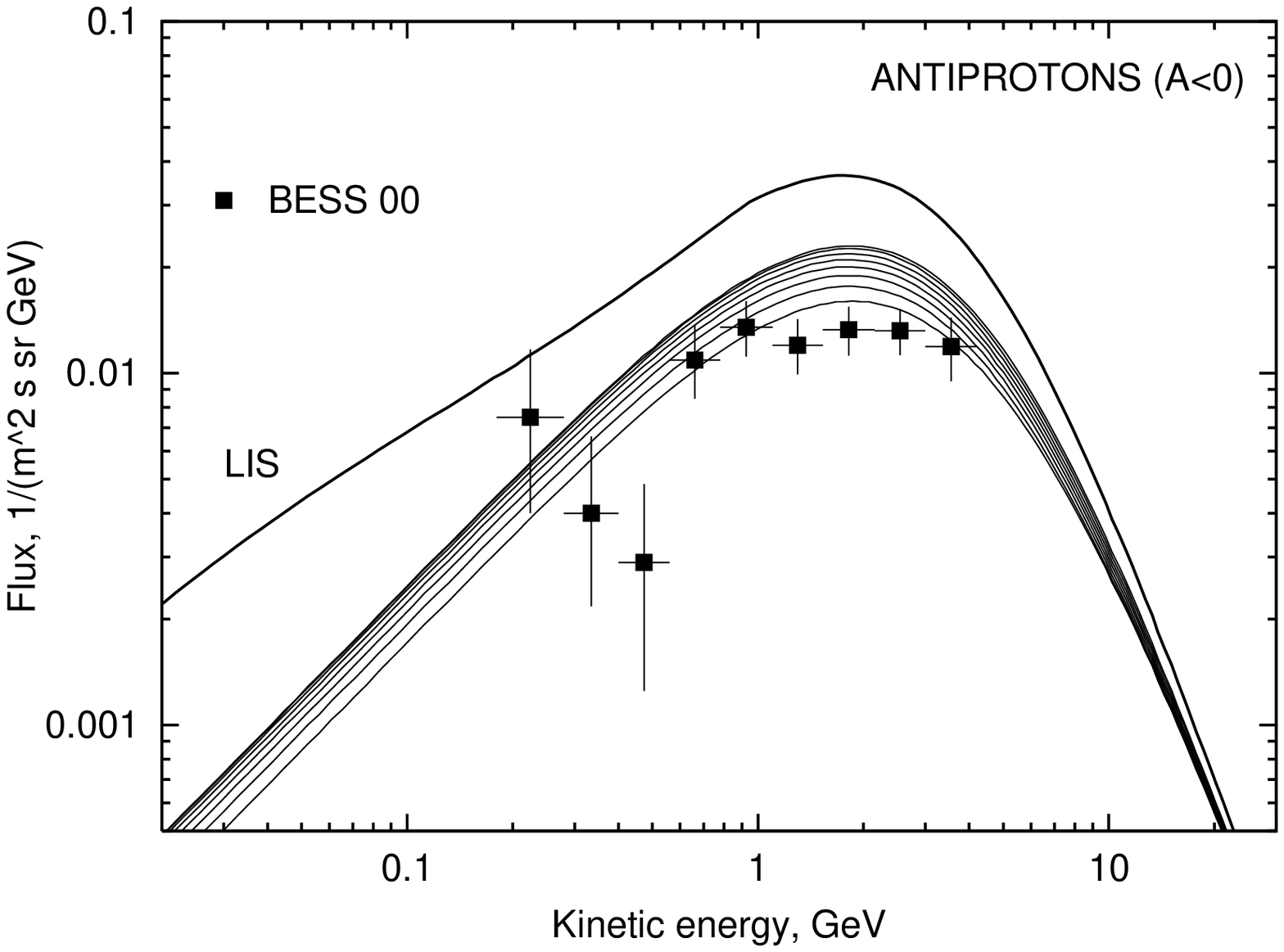,width=\onefig,clip=}}
\figcaption[f9a.ps,f9b.ps]{ Calculated antiproton LIS and modulated
spectra for the two magnetic polarity dependent modulation epochs,
$A>0$ (top) and $A<0$ (bottom). Tilt angle from top to bottom:
$5^\circ, 15^\circ, 25^\circ, 35^\circ, 45^\circ, 55^\circ, 65^\circ,
75^\circ$. The tilt angle corresponding to BESS data is
$\sim5^\circ-15^\circ$ ($A>0$) depending on the coronal field model.
Data references as in Figure \ref{fig:pbars}, BESS-99, BESS-00 \citep{asaoka01}.
\label{fig:pbar_cycle} \bigskip}
\end{figure*}

\begin{figure*}[!p]
\centerline{\psfig{file=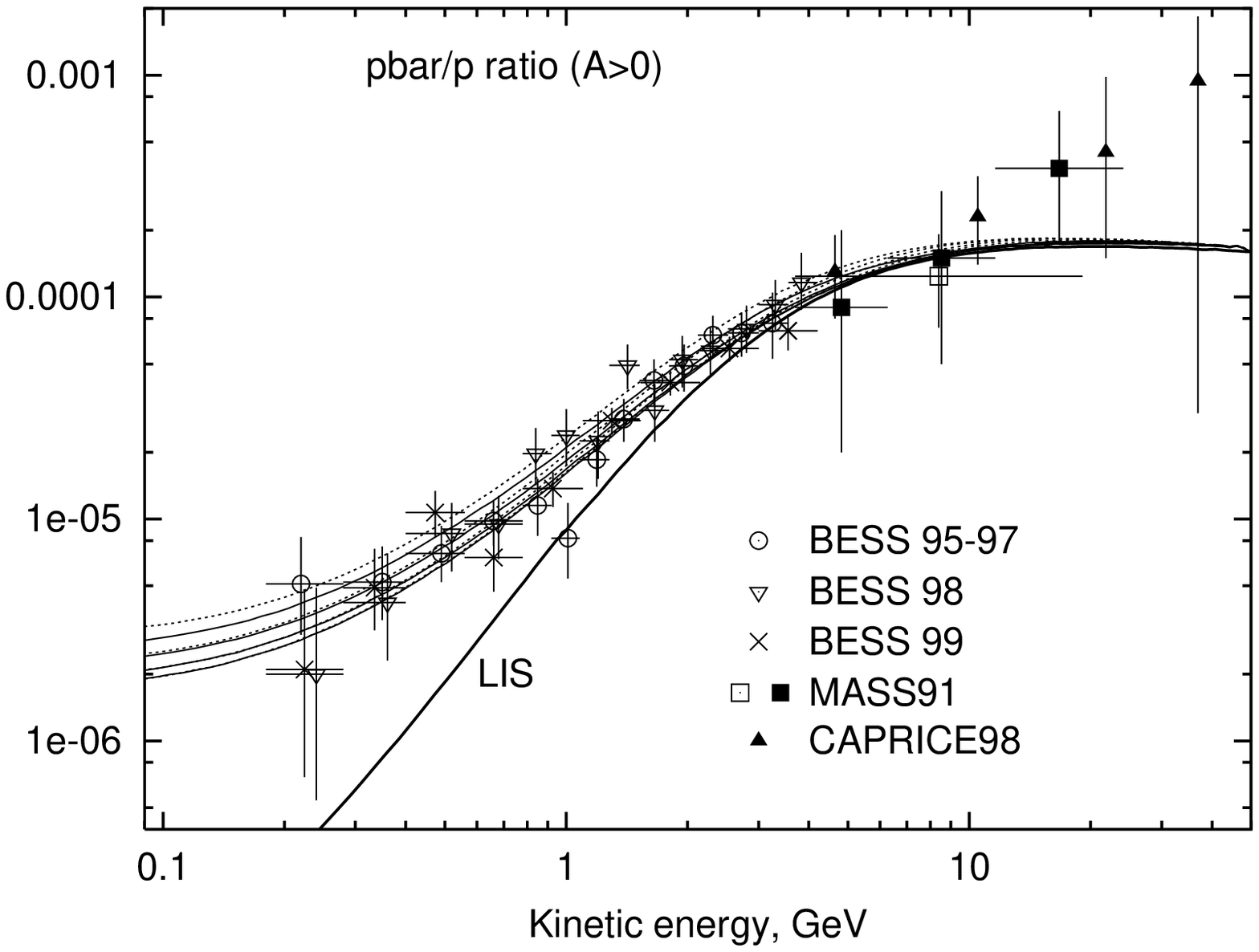,width=\onefig,clip=}}
\centerline{\psfig{file=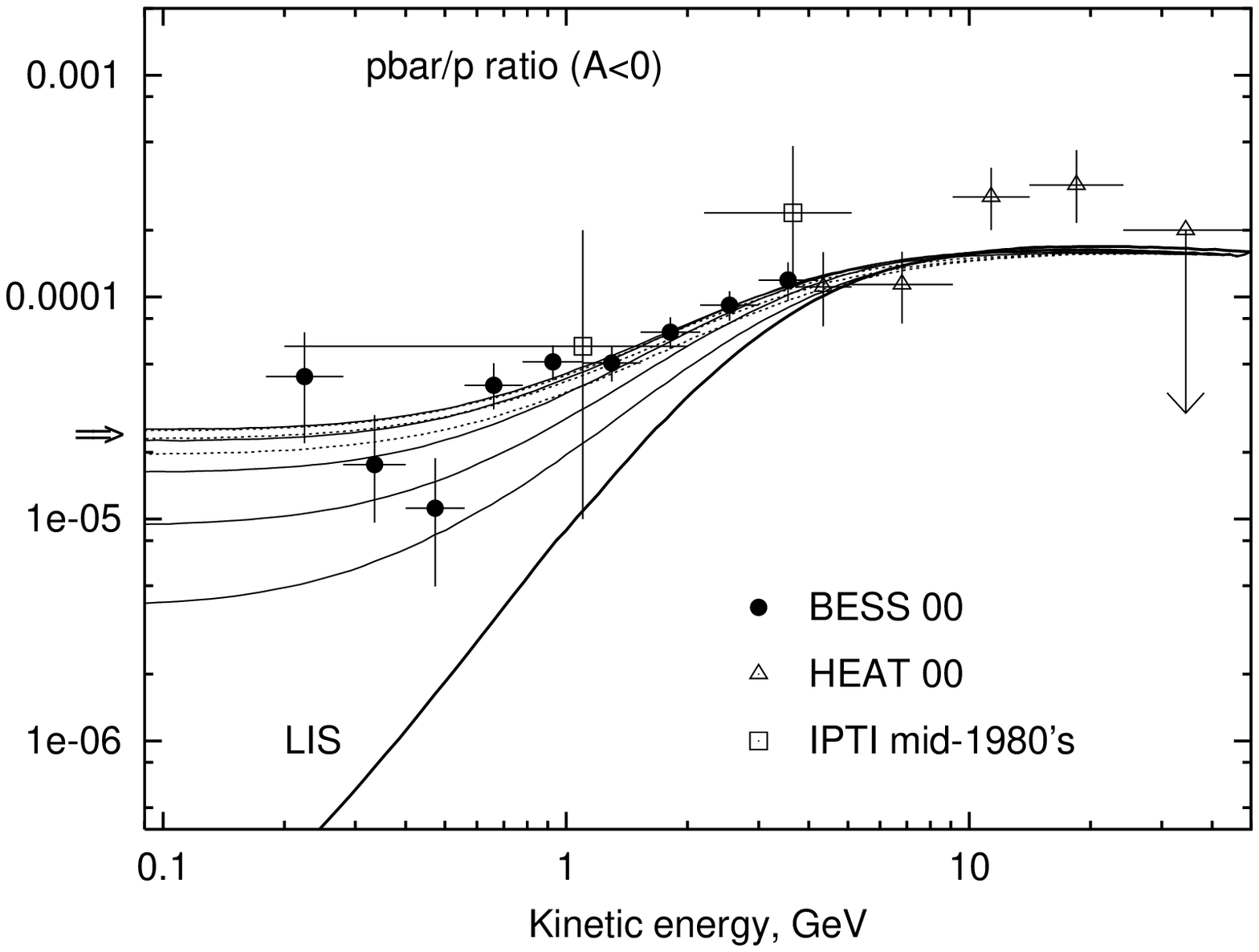,width=\onefig,clip=}}
\figcaption[f10a.ps,f10b.ps]{ Calculated $\bar p/p$ ratio in the
interstellar medium (LIS) and modulated for the two magnetic polarity
dependent modulation epochs, $A>0$ and $A<0$. Top:
Tilt angle from \emph{top to bottom} (solid lines): $5^\circ,
15^\circ, 25^\circ, 35^\circ$; from \emph{top to bottom} (dotted
lines): $75^\circ, 65^\circ, 55^\circ, 45^\circ$. (Lines almost
coincide for the tilt angles $35^\circ, 45^\circ$ and $25^\circ,
55^\circ$, and very close for $15^\circ, 65^\circ$.) 
Data: BESS \citep{Orito00,Maeno01,asaoka01}, MASS91 \citep{hof,Stoc01}, and CAPRICE98
\citep{bergstroem00}.
Bottom: Tilt
angle from \emph{bottom to top} (solid lines): $5^\circ, 15^\circ,
25^\circ, 35^\circ, 45^\circ$; from \emph{bottom to top} (dotted
lines): $75^\circ, 65^\circ, 55^\circ$. (Lines are very close for the
tilt angles $35^\circ, 65^\circ$ and $45^\circ, 55^\circ$.)
The BESS-00 data agree well with our predictions for 
the tilt angles $55^\circ-65^\circ$ (indicated by the arrow). 
Data: IPTI \citep{bogomolov2,bogomolov3}, HEAT-00 \citep{heat2000},
BESS-00 \citep{asaoka01}.
\label{fig:pbar2p} \bigskip}
\end{figure*}

We use the DC model to calculate the LIS spectra of protons and
antiprotons and then use the drift model to determine their modulated
spectra and ratio over the solar cycles with positive ($A>0$) and
negative ($A<0$) polarity (Figs.\
\ref{fig:prot_cycle}--\ref{fig:pbar2p}). The variations shown depend
on the tilt angle. When combined with Figure \ref{fig:tilt} this
allows us to estimate the near-Earth spectra for arbitrary epochs in
the past as well as make some predictions for the future. It may be
also used to test the theory of heliospheric modulation.

\placefigure{fig:prot_cycle}

\placefigure{fig:pbar_cycle}

\placefigure{fig:pbar2p}

The $\bar p/p$ ratio is supposed to be more accurately measured than
the flux. Interestingly, the $\bar p/p$ ratio appears largely
insensitive to the modulation level during the $A>0$ cycle (Fig.\
\ref{fig:pbar2p}, left). This agrees within the error bars with BESS
$\bar p/p$ measurements made in 1993, 1995, 1997, 1998
\citep{Maeno01}, and 1999 \citep{asaoka01}, and
thus allows us to constrain the proton spectrum
at low energies, in the heliosphere above $\sim0.1$ GeV and the LIS
spectrum above $\sim0.7$ GeV. In the new cycle ($A<0$), on the
contrary, this ratio is predicted to vary by over an order of
magnitude (Fig.\ \ref{fig:pbar2p}, right), which will allow to us
validate the calculations of heliospheric modulation including charge
sign effects.
Our predictions for tilt angles $55^\circ-65^\circ$ (indicated by the arrow) 
agree well with the data from the BESS-2000 flight \citep{asaoka01}.
The HEAT-2000 data \citep{heat2000} were obtained during the Spring 2000 flight 
at around the solar maximum, but the solar modulation is weak at high energies.
Shown are also two data points obtained at the moderate level of solar activity 
in 1984--1985 and 1986--1989 in the previous $A<0$ cycle 
\citep[and Bogomolov, private communication]{bogomolov2,bogomolov3}. 

The proton flux data obtained during BESS-1999, 2000 flights \citep{asaoka01}
agree qualitatively with variations over the solar cycle.
However, BESS-1999 data show a flux which is somewhat larger
while BESS-2000 data show the flux lower than predicted. 
Both measurements have been made during maximum solar activity
and near the solar magnetic field reversal and may reflect the dynamic effects
taking place in the heliosphere.

Now we use the predicted variations over the solar cycle to combine
BESS data collected in 1995, 1997 during the solar minimum with BESS
data collected in 1998, when the solar activity was moderate (the tilt
angle about $25^\circ$). The correction was calculated using the
energy-dependent factor, $F(E)_{5/25}=\psi_{5}/\psi_{25}$, the ratio
of modulated antiproton spectra with tilt angles $5^\circ$ and
$25^\circ$. The data of 1998 and their error bars have been
multiplied by this factor. The spectrum obtained in such a way has
been further combined with BESS data collected in 1995, 1997. 

\placefigure{fig:pbar_combined}

The combined spectrum agrees very well with calculations for tilt
angle $5^\circ$ and $A>0$ (Fig.\ \ref{fig:pbar_combined}).
It is, however, not sensitive to the exact value of
the tilt angle at the epoch of BESS-98 flight. It looks very similar
for tilt angles $15^\circ$ and $35^\circ$. The reason is that the
error bars of the data collected in 1998 are large, so that their
relative weight is small.

\section{Conclusions} \label{sec:concl}

In this paper, we have studied five basic propagation models. Table
\ref{table5} summarizes the results of this study. First, we applied a
propagation model with reacceleration. It reproduces nuclear
secondary/primary ratios rather well but has some difficulties with
reproduction of primary proton and He spectra, and, more important
with secondary positrons and antiprotons. 
When the primary injection spectra are adjusted to fit the
observed modulated $p$ and He spectra, the secondary positron
excess is reduced but still significant, while the antiproton
deficit is unaffected.
Another model with reduced
reacceleration strength and convection also produces too few
antiprotons. A plain diffusion model (no reacceleration, no
convection) reproduces the high energy part of nuclear
secondary/primary ratios, protons, and positrons, but it has a problem
with the low energy part of nuclear secondary/primary ratios and
overestimates the antiproton flux. A model with convection and
flattening of the diffusion coefficient at low energies and a break
in the injection spectrum is our ``best-fit'' model. It reproduces
well all particle data ``on average''. The spectrum of diffuse \grays\
calculated for this model, cannot explain the EGRET GeV excess
\citep{hunter97}, but this could originate in other ways.

\placetable{table5}

During the last decade there have been a number of space and balloon
experiments with improved sensivity and statistics. They impose
stricter constraints on the CR propagation models. It has become
clear that currently there is no \emph{simple} model that is
able to simultaneously reproduce all data related to CR origin and
propagation. This conclusion is mainly the result of the
increased precision of the CR experimental data, but also the improved
reliability of the calculations of interstellar and heliospheric 
propagation.


\begin{table*}[!tbh]
\tablecolumns{7}
\tablewidth{0mm}
\tabletypesize{\footnotesize}

\caption{Summary of model predictions. \label{table5}}
\begin{tabular}{lcccccc}\hline\hline\noalign{\smallskip}

\colhead{} & 
\colhead{} & 
\multicolumn{2}{c}{Primaries} & 
\colhead{} & 
\multicolumn{2}{c}{Secondaries}
\\
\cline{3-4}\cline{6-7}
\colhead{Model} & 
\colhead{B/C} &
\colhead{Protons} & 
\colhead{He} &
\colhead{} & 
\colhead{Antiprotons} & 
\colhead{Positrons}
\startdata
DR\tablenotemark{a}  & 
Good                 &
LE bump/Good         &
Fair/Good            & 
 &
Too few              &
LE bump              \\

DRB                  & 
Good                 &
Good                 &
Good                 & 
 &
Too few              &
LE excess            \\

DR\tablenotemark{a}  & 
Good                 &
LE bump/Good         &
Fair/Good            & 
 &
Too few              &
LE bump              \\

MRC\tablenotemark{a} &
Fair                 &
LE bump/Good         &
Fair/Good            & 
 &
Too few              &
LE bump              \\

PD                   &
Too large at LE      &
Good                 &
Good                 & 
 &
Too many             &
Good                 \\


DC                   &
Fair                 &
Good                 &
Good                 & 
 &
Good                 &
Good                 
\\
\enddata
\tablenotetext{a}{The reproduction of spectra of primaries can be improved 
by choice of the injection spectrum.}

\end{table*}

What could be the origin of this failure (to find a simple model), 
apart from the propagation models?
Concerning the accuracy of the experimental data, the spectra of
protons and helium are measured almost simultaneously and quite
precisely by BESS and AMS and the agreement is impressive. They also
agree with earlier experiments, such as LEAP, IMAX, CAPRICE, within
the error bars. The most accurate measurements of nuclei at low
energies are made by Voyager, Ulysses, and ACE, and the agreement is
good. At higher energies the data obtained by HEAO-3 are the most
accurate and generally agree with earlier measurements. Electron flux
measurements made by HEAT, CAPRICE, and at Sanriku balloon facility
all agree. Positron data, though with large error bars, are in
agreement as well. The possibility that BESS antiproton data have
systematic errors as large as +20\% looks improbable. They have
carried out calibrations which demonstrate that the error
must be $\le5$\% \citep{bess_calibration}. However,
additional measurements are desirable.

\centerline{\psfig{file=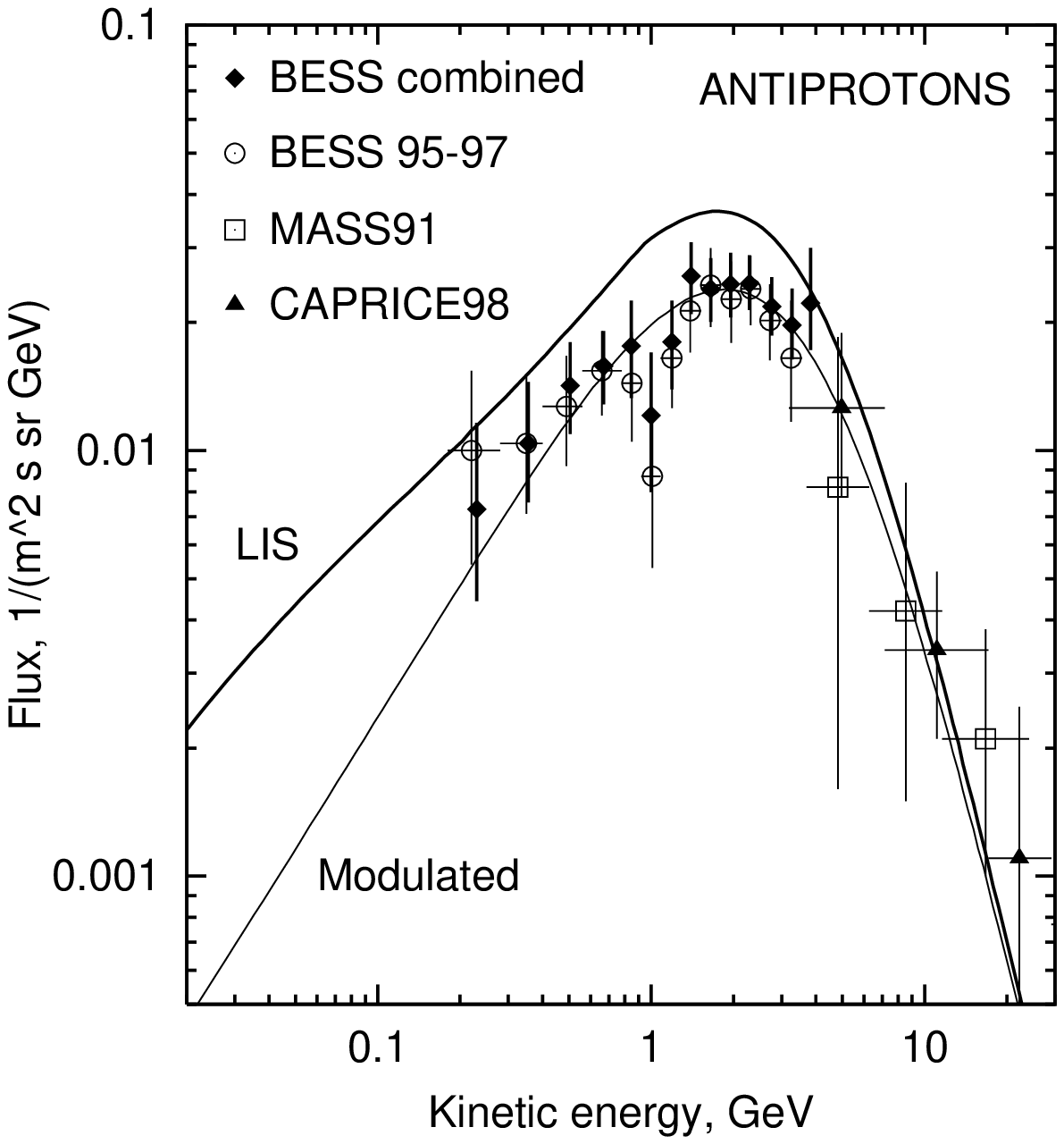,width=\twofigs,clip=}}
\figcaption[f11.ps]{ BESS 1995-97 data and the data combined with
``corrected'' BESS 1998 measurements. Shown also are the calculated
antiproton interstellar (LIS) and modulated spectrum for a tilt angle
$5^\circ$ ($A>0$). Data references as in Figure \ref{fig:pbars}.
\label{fig:pbar_combined} \bigskip}

Nuclear cross section errors are one of the main concerns. Fitting
(matching) the measured B/C ratio is a standard procedure to derive
the propagation parameters. The calculated ratio, in turn, depends on
many cross sections, such as the total interaction and isotope
production cross sections. The latter appear to have quite large
uncertainties, typically about 20\%, and sometimes they can even be
wrong by an order of magnitude \citep[see, e.g.,][]{MMS01}. This is
reflected in the value of the diffusion coefficient, the Alfv\'en
velocity (parameterizing reacceleration) and/or convection velocity,
and thus influences the calculated spectra of CR species. Our cross
section calculations make use of the fits to the cross sections
$p+\mathrm{ C,N,O} \to \mathrm{ Be,B}$, that produce most of the Be and
B. Other channels are calculated using the \citeauthor{W-code} or
\citeauthor{ST-code} cross section codes renormalized to data where it
exists. We thus can rule out a possibility of large errors in the
calculated B/C ratio.

Solar modulation for antiprotons is different from that of protons,
the effect known as charge sign dependence. Over the last years
Ulysses made its measurements at different heliolatitudes so we know
more about the solar magnetic field configuration and the solar wind
velocity distribution. On the other hand, the two Voyagers, the most
distant spacecraft, provide us with measurements of particle fluxes
made close to the heliospheric boundary. There is thus a small chance
of a serious error, e.g., that the antiproton flux during the last
solar minimum was modulated much more weakly than estimated.

The underproduction of antiprotons in the reacceleration model may be
connected with a contribution of primary antiprotons, but this
suggestion conflicts with other CR data. A strong primary
antiproton signal would be accompanied by the positron signal and an
excess in \grays\ (EGRET GeV excess ?). However, the positron flux
calculated in the DR model already shows an excess rather than deficit.
Besides, only few SUSY dark matter candidates are able to produce a
signal large enough to be detected and even in this case primary
antiprotons contribute mostly to low energies \citep[see discussions
in][]{bottino98,berg99}.

Considering reacceleration models, there is the possibility
that the injection spectra of protons and nuclei might be
different. For instance, spectra of nuclei $Z>2$ are well reproduced
by a reacceleration model with a power-law (in rigidity) injection
spectrum, while protons, helium, and electrons require some spectral
flattening at low energies to avoid developing a bump in the spectrum.
This is in agreement with conclusions made by other authors
\citep{jones01}. The difficulty is to get the right positron and
antiproton fluxes in the reacceleration model. On the other hand,
non-reacceleration models require some spectral steepening at low
energies (compensation for flattening of the diffusion coefficient) 
to get good agreement with
the spectra of primaries. Such steepening may be a consequence of SNR
shock acceleration \citep{ellison}.

If we assume that the problem is in propagation models, we have shown
that it \emph{is} possible to construct a model (DC) that fits all
these data, by postulating a significant flattening of the diffusion
coefficient below 4 GV together with convection (and possibly with
some reacceleration) and a break in the injection spectrum. The
break in the diffusion coefficient is reminiscent of the standard
procedure in ``leaky-box'' models where the escape time is set to a
constant below a few GeV. This has always appeared a completely
\emph{ad hoc} device without physical justification, but the present
analysis suggest it may be forced on us by the data. Therefore,
possibilities for its physical origin should be studied.

Where do we go from here? New measurements of CR species that cover
the range $0.5-1000$ GeV/nucleon are necessary to distinguish between
reacceleration and non-reacceleration models. The new experiment
PAMELA scheduled for launch in 2002 should improve accuracy of
positron (0.1--200 GeV) and electron (0.1--300 GeV) measurements while
allowing for simultaneous measurements of $\bar p$ and nuclei from H
to C in the energy range 0.1--200 GeV/nucleon \citep{pamela}. The
puzzling excess in the diffuse \grays\ above 1 GeV in EGRET data
should be confirmed. Here we await the launch of GLAST with 30 times
better sensivity. Its improved angular resolution will be helpful in
removing the unresolved source component contribution to the diffuse flux. A
possibility that the nucleon spectrum in the Galaxy is harder or
softer than we measure locally should be explored by making more
accurate measurements of antiproton fluxes at high energies
\citep{MSR98}. This may help to produce more antiprotons in the DR 
model and simultaneously explain the GeV excess in \grays. The
interpretation of B/C and other secondary/primary ratios depends on
the assumption that all the secondaries are produced during
propagation; if instead some secondary production occurs in the
sources the range of possible models widens. Whether such effects
could provide an alternative explanation of the observed effects
should be the subject of future study. 
The effect of the local distribution of interstellar matter is worth
investigating and we are going to do this in future by developing and
incorporating in our code a 3D model of the gas distribution in the
Galaxy.  The effect should however only important for radioactive  and
K-capture isotopes and electrons/positrons, i.e.\ for those particles
whose half-life is less or comparable to the cosmic ray life-time in
the Galaxy or whose large energy losses confine them to a short
distance from the source. On the other hand for protons, with their 30
mb total cross section and negligible energy losses, only the
large-scale gas distribution is important. The same is true for
antiprotons, as their total interaction cross section at 1 GeV is only
about two times larger and approaches the proton cross  section at
higher energies (Figure \ref{fig:cs_plots}).
A thorough and systematic
measurement of the nuclear cross sections over a wide energy range is
extremely important. Accurate measurements of the cross sections would
allow more information to be extracted from the now rather precise CR
measurements. New measurements of the antiproton and charged and
neutral pion production cross sections are desirable. The widely used
parametrization of antiproton production and interaction cross
sections by \citet{TanNg83a,TanNg83b} is twenty years old.
Parametrization of pion production is even older. Given the new
experimental techniques now available such measurements should be
straightforward.
At the same time the sensitivity of the whole
analysis to the modulation models has to be investigated further.

\acknowledgments

The authors are grateful to
Frank Jones and Vladimir Ptuskin for fruitful discussions, to Don
Ellison for discussions and a copy of his manuscript prior to
publication, to S.\ A.\ Stephens and R.\ A.\ Streitmatter for
providing the database of CR measurements, and to Ulrich Langner for his
interest and assistance in producing modulation runs for Figures
\ref{fig:prot_cycle} to \ref{fig:pbar2p}. I.\ Moskalenko acknowledges
support from NAS/NRC Research Associateship Program.

\appendix
\section{The interstellar gas density distribution in the cylindrically
symmetrical model} \label{gas}

The H$_2$ number density in mols.~cm$^{-3}$ is calculated from
\begin{equation}
\label{eq.18}
n_{\mathrm{H}_2}(R,z) = 3.24\times10^{-22} X \epsilon_0(R)\, e^{-\ln2\ (z-z_0)^2/z_h^2}, 
\end{equation}
where $\epsilon_0(R)$ (K km s$^{-1}$ kpc$^{-1}$) is the CO volume
emissivity, $z_0(R)$ and $z_h(R)$ are the height scale and width
defined by a table \citep{Bronfman88}, and $X \equiv
n_{\mathrm{H}_2}/\epsilon_{\mathrm{CO}}=1.9\times 10^{20}$
mols.~cm$^{-2}$/(K km s$^{-1}$) is the conversion factor
\citep{StrongMattox96}.

The \hi\ (atom cm$^{-3}$) relative distribution is taken from 
\citet{GordonBurton76}, but renormalized to agree with \citet{DL90},
since they give their best model for
the $z$-distribution in the range $R=4-8$ kpc and state the total
integral perpendicular to the plane is $6.2\times10^{20}$ cm$^{-2}$:
\begin{equation}
\label{eq.19}
n_\mathrm{H\ I}(R,z) = \frac1{n_\mathrm{GB}}Y(R)\left\{
\begin{array}{ll}
\sum_{i=1,2} A_i\, e^{-\ln2\ z^2/z_i^2} +A_3\, e^{-|z|/z_3},
   & R\leq 8\ \mathrm{kpc}\\
\mathrm{interpolated,} 
   & 8\!<\! R \!<\!10\ \mathrm{kpc} \\
n_\mathrm{DL} \exp(- z^2\,e^{-0.22R}/z_4^2),
   & R\geq 10\ \mathrm{kpc}
\end{array} \right.
\end{equation}
Here $Y(R)$ is the distribution from \citet{GordonBurton76}
($R<16$ kpc), $n_{\mathrm{GB}}= 0.33$ cm$^{-3}$ and
$n_{\mathrm{DL}}=0.57$ cm$^{-3}$ are the disk densities in the range
$4<R<8$ kpc in models by \citet{GordonBurton76} and \citet{DL90},
correspondingly. The $z$-dependence is calculated using the approximation
by \citet{DL90} for $R< 8$ kpc, using the approximation by
\citet{Cox86} for $R>10$ kpc, and interpolated in between,
and the parameter values are $A_1 =0.395$, 
$A_2 =0.107$, $A_3 =0.064$, $z_1 =0.106$, $z_2 =0.265$, $z_3 =0.403$, $z_4
=0.0523$. For $R>16$ kpc an exponential tail is assumed with scale
length 3 kpc.

The ionized component \hii\ (atom cm$^{-3}$) is calculated using a
cylindrically symmetrical model \citep{Cordes91}:
\begin{equation}
\label{eq.20}
n_\mathrm{H\ II}(R,z) =
\sum_{i=1,2} n_i\, e^{-|z|/h_i -(R-R_i)^2/a_i^2},
\end{equation}
where $n_1 =0.025$, $n_2 =0.200$, $h_1 =1$ kpc, $h_2 =0.15$ kpc,
$R_1=0$, $R_2 =4$ kpc, $a_1 =20$ kpc, $a_2 =2$ kpc.

\section{Proton and antiproton cross sections} \label{sec:appendix}

The energy and momentum units are GeV and GeV/c correspondingly.
All cross section given are in mbarn and plotted in 
Figure \ref{fig:cs_plots} (except for the production cross section).
An asterisk marks the center-of-mass system (CMS) variables.

\placefigure{fig:cs_plots}

\centerline{\psfig{file=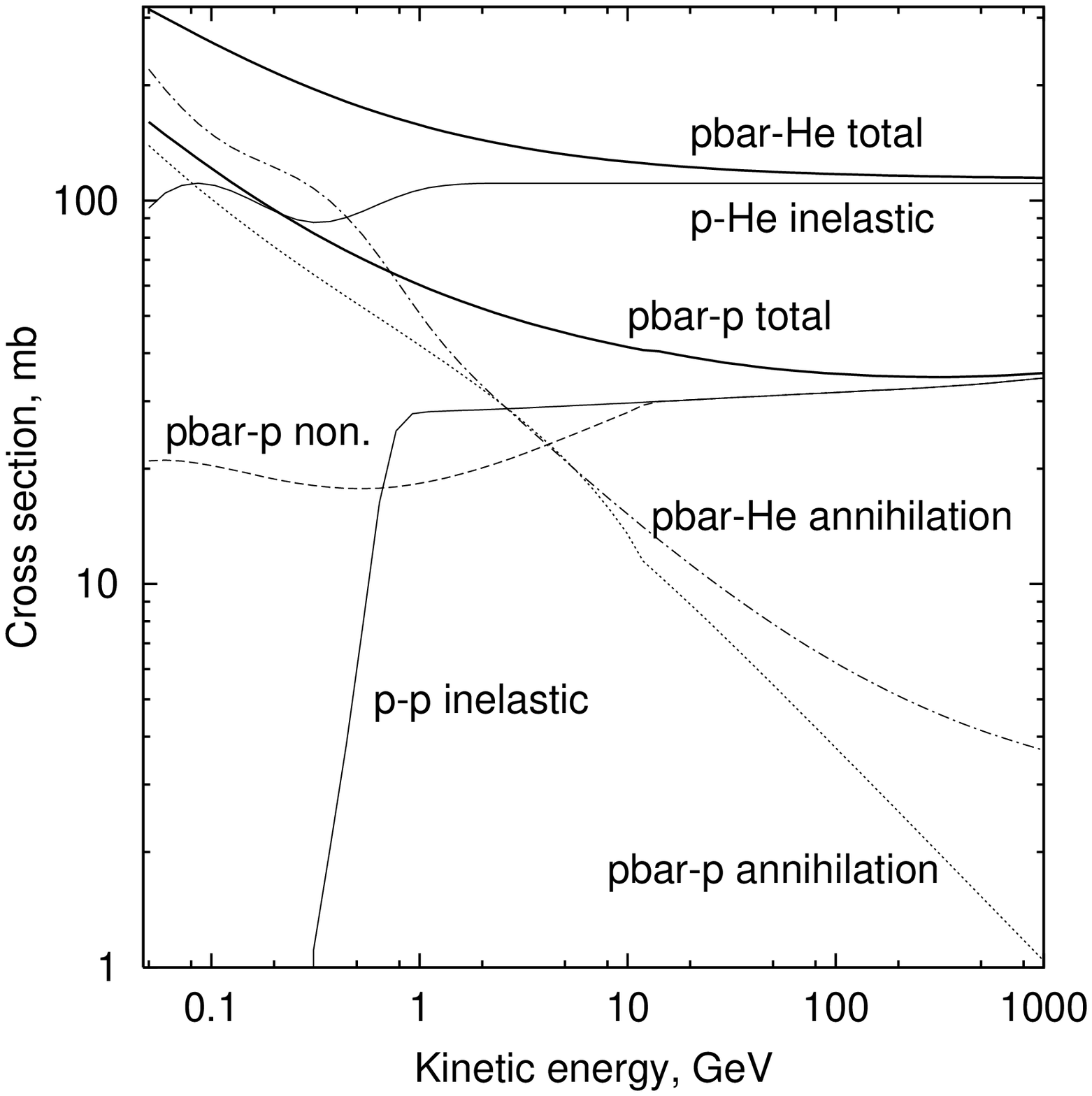,width=\twofigs,clip=}}
\figcaption[f12.ps]{ Proton and antiproton cross sections.
\label{fig:cs_plots} \bigskip}

The inclusive antiproton production cross section (mb GeV$^{-2}$ c$^3$)
in $pp$-reaction is given by \citep{TanNg83b}: 
\begin{eqnarray}
\label{A.01}
E_{\bar p}\frac{d^3\sigma}{dp_{\bar p}^3}\ &=& 
       \delta(x_t) f(x_r) \exp\{-A(x_r)p_t+B(x_r)p_t^2\},\\ 
f  &=& 1.05\times10^{-4} \exp(-10.1 x_r)\,\theta(0.5-x_r) +3.15(1-x_r)^{7.9},\nonumber \\
A  &=& 0.465\,\exp(-0.037 x_r) +2.31\,\exp(0.014 x_r),\nonumber \\
B  &=& 0.0302\,\exp\{-3.19 (x_r+0.399)\}(x_r+0.399)^{8.39},\nonumber\\
x_r&=& E_{\bar p}^*/  E_{\bar p}^{*\, max},\nonumber
\end{eqnarray}
where $\theta(x)$ is the Heaviside step function ($\theta[x>0]=1$, otherwise =0),
$p_t$ is the transverse momentum of the antiproton,
$E_{\bar p}^*$ is the total CMS energy of the antiproton, 
$E_{\bar p}^{*\, max}$ is the maximal value of $E^*_{\bar p}$ for the given
inclusive reaction, and
$\delta(x_t)$ is the low energy correction, $\delta=1$ at $s^{1/2} > 10$ GeV.
At $s^{1/2} \leq 10$ GeV it is given by:
\begin{eqnarray}
\label{A.02}
\delta^{-1}&= &1-\exp\left\{-\exp\left[c(x_t)Q-d(x_t)\right]
   \left(1-\exp\left[-a(x_t)Q^{b(x_t)}\right]\right)\right\},\nonumber\\
a  &=& 0.306\, \exp(-0.12 x_t),\nonumber\\
b  &=& 0.0552\, \exp(2.72 x_t),\nonumber\\
c  &=& 0.758\, -0.68 x_t +1.54 x_t^2,\\
d  &=& 0.594\, \exp(2.87 x_t),\nonumber\\
Q  &=& \sqrt{s}-4M_p,\nonumber\\
x_t&=& T_{\bar p}^*/T_{\bar p}^{*\,max},\nonumber
\end{eqnarray}
$s=2 M_p (E_{\bar p}+M_p)=inv$ is the square of the total CMS energy of colliding
particles, $M_p$
is the proton rest mass, $T_{\bar p}^*$ is the kinetic CMS energy of antiproton, and
$T_{\bar p}^{*\, max}$ is the maximal value of $T_{\bar p}^*$ for the given
inclusive reaction.

Proton-proton inelastic cross section \citep{TanNg83a}:
\begin{eqnarray}
\label{A.1}
\sigma_{pp}^{inel} &=& 32.2\,[1+0.0273 U+0.01 U^2\theta(U)]\left\{
\begin{array}{ll}
0,                                              &          T_p < 0.3;\\
\left(1+2.62\times10^{-3} T_p^{-C}\right)^{-1}, & 0.3 \leq T_p < 3;\\
1,                                              &          T_p \geq 3; 
\end{array} \right.\nonumber \\
U &=& \ln (E_p/200),\nonumber \\
C &=& 17.9 +13.8\ln T_p +4.41\ln^2 T_p,\nonumber 
\end{eqnarray}
where $\theta(U)$ is the Heaviside step function, $E_p$ and $T_p$
are the total and kinetic energy of proton, correspondingly.

Nucleus-proton inelastic cross section \citep{Letaw83}:
\begin{eqnarray}
\label{A.2}
\sigma_{pA}^{inel} &=&45 A^{0.7} \delta(T_p)\left[1+0.016 \sin(5.3-2.63\ln A)\right]
\left\{
\begin{array}{ll}
1-0.62 e^{-T_p/0.2} \sin\left[\frac{10.9}{(10^3T_p)^{0.28}}\right], & T_p \leq 3;\\
1,                                                                  & T_p > 3;
\end{array} \right.\nonumber\\
\delta &=& \left\{
\begin{array}{ll}
1+0.75 \exp\{-T_p/0.075\}, & \mathrm{for\ beryllium};\\
1,                         & \mathrm{otherwise};
\end{array} \right.\nonumber
\end{eqnarray}
where $A$ is the atomic number.
In the case of $p$He inelastic cross section equation (\ref{A.2}) is known to be
not very accurate, while the $p-^4$He cross section is the most important 
after the $pp$ cross section. We thus made our own fit to the data in the range
$0.02-50$ GeV:
\begin{equation}
\label{A.2.1}
\sigma_{pHe}^{inel}= 111\left\{1-\exp [-3.84\,(T_p-0.1)]
\left(1-\sin[9.72\log^{0.319}(10^3T_p)-4.14]\right)\right\},
\end{equation}
where $T_p>0.01$ GeV.

The antiproton-proton annihilation cross section at low energies is taken 
from \citet{TanNg83a}. At high energies it is calculated as 
the difference between the total $\bar pp$ and $pp$ cross sections
which are parametrized using Regge theory \citep{PDG}:
\begin{equation}
\label{A.3}
\sigma_{\bar pp}^a = \left\{
\begin{array}{ll}
661\left(1 +0.0115\,T_{\bar p}^{-0.774}  -0.948\,T_{\bar p}^{0.0151}\right),
                        & T_{\bar p}\leq10;\\
2\times 35.43s^{-0.56} ,& T_{\bar p}>10;\\
\end{array} \right.
\end{equation}
where $s$ is the square of the total CMS energy of colliding
particles as defined in equation (\ref{A.02}). 
Annihilation is of minor importance at high energies.

Total {\it inelastic} antiproton-proton cross section \citep{TanNg83a}:
\begin{equation}
\label{A.4}
\sigma_{\bar pp}^{tot} = \left\{
\begin{array}{ll}
24.7\left(1+0.584\,T_{\bar p}^{-0.115}
   +0.856\,T_{\bar p}^{-0.566}\right),      & T_{\bar p} \leq 14;\\
\sigma_{pp}^{inel} +\sigma_{\bar pp}^a,     & T_{\bar p} > 14.
\end{array} \right.
\end{equation}
At very low energies, below $\sim0.01$ GeV, 
we put $\sigma_{\bar pp}^a=\sigma_{\bar pp}^{tot}$.

Non-annihilation inelastic $\bar pp$ cross section is calculated as
$\sigma_{\bar pp}^{non} = \sigma_{\bar pp}^{tot}-\sigma_{\bar pp}^a$.
The antiproton-nucleus non-annihilation inelastic cross section 
is taken equal to proton-nucleus inelastic cross section,
$\sigma_{\bar pA}^{non} = \sigma_{pA}^{inel}$.

The parametrization of the
$\bar p$ total inelastic cross section on an arbitrary nuclear target
has been obtained in \citet{pbar_A} using a parametrization by
\citet*{kuzichev}
and has been tested against the data available on $\bar p$ cross sections
on C, Al, and Cu targets:
\begin{equation}
\label{A.5}
\sigma_{\bar pA}^{tot} = A^{2/3}\left[48.2+19\,T_{\bar p}^{-0.55}
+(0.1-0.18\,T_{\bar p}^{-1.2})Z+0.0012\,T_{\bar p}^{-1.5} Z^2\right],
\end{equation}
where $Z$ is the nucleus charge.
The second term in square brackets has been modified from its original
form $19\,(T_{\bar p}-0.02)^{-0.55}$ to better match the slope
below $\sim100$ MeV. This modification does not affect agreement with
data at higher energies. Another modification, 
in the case of $^4$He, is to use $A=3.3$ to scale the value down slightly
to be consistent with cross section data both at low and at
high energies \citep{pbar_He1,pbar_He2}.

The $\bar pA$ annihilation cross section is calculated from 
$\sigma_{\bar pA}^a=\sigma_{\bar pA}^{tot}-\sigma_{\bar pA}^{non}$, 
where $\sigma_{\bar pA}^{non} = \sigma_{pA}^{inel}$. This is
quite accurate provided that the He abundance is about 10\% that of H. 


\end{document}